\documentclass[fleqn,usenatbib]{mnras}

\usepackage{newtxtext,newtxmath}

\usepackage[T1]{fontenc}
\usepackage{pdflscape}

\DeclareRobustCommand{\VAN}[3]{#2}
\let\VANthebibliography\thebibliography
\def\thebibliography{\DeclareRobustCommand{\VAN}[3]{##3}\VANthebibliography}

\usepackage{graphicx}	
\usepackage{amsmath}	

\title[VLBA Survey of the Orion Nebula Cluster III]{\LARGE{A VLBA survey of radio stars in the Orion Nebula Cluster: III. Inter-epoch variability analysis}}

\author[O'Kelly et al.]{
Eoin T. O'Kelly$^{1}$,\thanks{E-mail: e.okelly@herts.ac.uk}
Jan Forbrich$^{1,2}$,
Sergio A. Dzib$^{3}$,
Jaime Vargas-Gonz$\acute{\text{a}}$lez$^{4}$
\\
$^{1}$Centre for Astrophysics Research, University of Hertfordshire, College Lane, Hatfield, AL10 9AB, UK\\
$^{2}$European Space Agency (ESA), European Space Research and Technology Centre (ESTEC), Keplerlaan 1, Noordwijk, 2201 AZ, The Netherlands. \\
$^{3}$Max-Planck-Institut f$\ddot{u}$r Radioastronomie, Auf dem H$\ddot{u}$gel 69, D-53121 Bonn, Germany\\
$^{4}$Joint ALMA Observatory, Alonso de Córdova 3107, Vitacura, Santiago, Chile
}

\date{Accepted XXX. Received YYY; in original form ZZZ}

\pubyear{2026}

\begin{document}
\label{firstpage}
\pagerange{\pageref{firstpage}--\pageref{lastpage}}
\maketitle


\begin{abstract}
We present a multi-epoch Very Long Baseline Array (VLBA) survey of radio sources in the Orion Nebula Cluster (ONC), updating the detection statistics and investigating inter-epoch variability. The dataset comprises eight C-band (4-8 GHz) epochs targeting 575 radio sources previously identified in deep Very Large Array surveys. 
VLBA observations result in resolutions of milliarcseconds and are only sensitive to high brightness temperatures meaning only non-thermal emission is observed. We detect 226 unique radio sources corresponding to a detection fraction of 39 per cent. The majority of sources ($\sim\!83$ per cent) are detected only once indicating that non-thermal emission associated with Young Stellar Objects (YSOs) is highly intermittent.
To quantify variability, we use per-epoch peak flux densities and the variability factor (VF) defined as the ratio between the maximum and minimum fluxes. We identify 21 sources ($\sim\!9$ per cent of detected sources) that exhibit high variability defined as a change of flux density by at least a factor of five.
We crossmatch with the COUP \textit{Chandra} survey to identify sources as YSOs. We identify COUP counterparts for $\sim\!56$ per cent of the detected sources. The remaining population cannot be explained solely by variability or background contamination, suggesting a population of radio detected YSOs without X-ray counterparts.
We find a clear correlation between radio and X-ray luminosities consistent with the G\"udel-Benz relation, the empirical relation between non-thermal radio and thermal X-ray emission for magnetically active stars.
\end{abstract}





\section{Introduction}
\label{sect:intro}

Star formation is associated with high-energy processes and Young Stellar Objects (YSOs) are known to be energetic radio sources \citep{Dulk:1985, Feigelson:1999}. Centimetric radio emission from YSOs can be either thermal free-free produced by ionised material or non-thermal (gyro)synchrotron generated by (mildly) relativistic electrons gyrating in magnetic fields associated with coronal-type activity \citep[e.g.,][]{Gudel:2002}.

High-energy events irradiating circumstellar disks may play an important role in disk evolution, chemistry and planet formation \citep[e.g.,][]{Feigelson:2002, Zhao:2020}. For example, \cite{Feigelson:2010} argues that high-energy irradiation may be critical in establishing disk turbulence via ionisation and enabling magneto-rotational instability, alongside other mechanisms such as hydrodynamical or gravitational instabilities.

Disentangling thermal free-free and non-thermal gyrosynchrotron emission can be challenging, with an interest in non-thermal emission as it traces high-energy processes associated with magnetic activity within the coronae of YSOs \citep[e.g.,][]{Feigelson:1999, Vargas-Gonzalez:2024-thesis}. The primary diagnostics used to distinguish between emission types are: polarisation (with gyrosynchrotron being circularly polarised and synchrotron linearly polarised), spectral index, variability and brightness temperature.

The spectral index, defined as $\alpha$ in $S_\nu\! \propto\! \nu^\alpha$, can provide insight into the emission mechanism where positive spectral indices are generally associated with thermal emission and negative indices with non-thermal emission \citep[e.g.,][]{Rybicki:1986, Wilson:2009, Scaife:2012}. However, interpretation of spectral indices can be ambiguous as there is a turnover within the spectrum and gyrosynchrotron emission can be time variable, these can correspond to different physical regimes \citep[e.g. optical depth, see][]{Dulk:1985, Benz:2002, Gudel:2002, Osten:2009}. Importantly, it should be highlighted that the relatively narrow frequency coverage of the observations presented here (see section \ref{sect:obs+data_reduction}) means in-band spectral indices would require higher signal-to-noise to reliably measure them \citep[e.g.,][discuss that uncertainties of measured spectral indices increase at lower signal-to-noise]{Rau:2011}.

Very Long Baseline Interferometry (VLBI) is sensitive only to emission with high brightness temperatures which effectively restricts detections to non-thermal processes \citep[e.g.,][]{Felli:1989, Thompson:2017}. Thermal free-free emission has a maximum brightness temperature $\lesssim\!10^5$K \citep[e.g.,][]{Condon:2016}, whereas non-thermal emission can reach temperatures of several million kelvin \citep[e.g.,][]{Andre:1996}. The minimum detectable brightness temperature can be expressed in terms of the flux density $S$, observing wavelength $\lambda$, and baseline length $D$, where $S=2kT_{B}\Omega/\lambda^{2}$ and the source solid angle is $\Omega \simeq \pi(\lambda/2D)^{2}$, giving the minimum detectable brightness temperature,

\begin{equation}
    (T_{B})_{\textbf{{min}}} \simeq \frac{2}{\pi k}D^{2}S_{min},
	\label{eq:TB}
\end{equation}

\noindent where $k$ is the Boltzmann constant \citep{Thompson:2017}. Equation \ref{eq:TB} shows that longer baseline length observations are only sensitive to high brightness temperatures.

The Very Long Baseline Array (VLBA) has a maximum baseline of 8611 km resulting in milliarcsecond resolutions. At these resolutions even a relatively faint detection of 0.2 mJy beam$^{-1}$ corresponds to a brightness temperature of $T_{B}\approx10^{7}K$ (Eq.\,\ref{eq:TB}), well above the maximum expected for thermal emission. Therefore, detections are associated with non-thermal emission. The smallest angular size that can be probed is limited by the synthesised beam size, if a source is unresolved the true angular size must be smaller than the size of the beam, therefore this corresponds to a lower limit of the brightness temperature \citep[e.g.,][]{Felli:1989, Andre:1991}. Consequently, VLBI is a powerful tool to isolate non-thermal radio emission from YSOs \citep[e.g.,][]{Forbrich2021, Dzib2021, Launhardt:2022, Dzib:2024}.

The Orion Nebula Cluster (ONC) is the nearest region of active high-mass star formation and provides a unique opportunity to observe a large number of sources within a single primary beam. At the observing frequency of this dataset the full width at half maximum (FWHM) of the primary beam is $\sim$7 arcmin, allowing simultaneous observations of many sources within the cluster. The ONC is relatively nearby, with recent distance estimates ranging from 414~$\pm$~7~pc \citep{Menten:2007}, 403$_{-6}^{+7}$ pc \citep{Kuhn:2019} to 388~$\pm$~5~pc \citep{Kounkel:2017} and the most recent of 388.5~$\pm$~1.7~pc \citep{Dzib:2026a}. When a single distance is required in this work we adopt the value from \cite{Dzib:2026a}.

The radio source population of the ONC has been studied extensively with the US National Radio Astronomy Observatory (NRAO) Very Large Array (VLA) since the earliest surveys by \cite{Garay:1987} and \cite{Churchwell:1987}, which detected 21 and 22 radio sources respectively. Subsequent VLA observations by \cite{Zapata:2004-1} expanded the radio population and detected 77 sources. A deep VLA survey conducted by \cite{Forbrich:2016} detected 556 sources. The source positions from this survey form the basis for the majority of the VLBA phase centres used in our work. In addition, 19 sources identified by the VLA survey of \cite{Vargas-gonz:2021} were also included resulting in a total of 575 targeted source positions. It is important to note that it is known that not all of these VLA detected sources are stellar in origin and would not be detected in VLBA observations \citep[e.g.,][both discuss detections of jets and outflows]{Forbrich:2016, Vargas-gonz:2021}.

Recent upgrades to both the VLA and VLBA have significantly improved sensitivity and observational capabilities. The development of the DiFX software correlator \citep{Deller:2007, Deller:2011} enables the correlation of many phase centres within a single primary beam required due to the field-of-view limitations inherent to VLBI observations (see section \ref{sect:data_analysis}). Whereas, earlier VLBI observations were limited to correlating only a small number of phase centres the DiFX software correlator allows all radio sources identified in the VLA surveys of the ONC discussed above to be targeted in a single observation.

Given that the VLBA observations presented here are sensitive exclusively to non-thermal radio emission the connection to X-ray studies of the ONC should be considered. An empirical correlation between non-thermal radio and thermal X-ray emission, the G\"udel-Benz relation, was first identified by \cite{Gudel-benz:1993} and \cite{Benz-gudel:1994}. This correlation has been further studied within the ONC using VLA \cite{Forbrich:2013} and by \cite{Yanza:2022} whose focus was on M17 (Omega Nebula or NGC 6618) but also studied this within the ONC.

For the X-ray perspective required here we use the \textit{Chandra} Orion Ultradeep Project (COUP) which provides a comprehensive X-ray census of the ONC based on observations with NASA's \textit{Chandra} X-ray Observatory \citep{Getman:2005}. This was also done for the other studies mentioned previously who used this X-ray census similarly. The COUP dataset consists of six nearly consecutive exposures spanning 13.2 days, a total exposure time of 9.7 days and covers an area of approximately 17 $\times$ 17 arcmin, which is significantly larger than the VLBA primary beam. The resulting catalogue contains over 1600 X-ray sources and is currently the most sensitive X-ray survey of the ONC. The COUP catalogue allows cross-matching with radio sources to identify them as being associated with the young stellar population, since elevated X-ray emission in this context is associated with magnetically active pre-main sequence stars \cite[e.g.,][]{Feigelson:1999}. However, the absence of an X-ray counterpart does not preclude it from being a YSO as extinction associated with deeply embedded objects and source variability can be factors \citep[e.g.,][]{Forbrich2021}. In this work, the COUP catalogue is used to investigate the empirical relationship between non-thermal radio emission and thermal X-ray emission, which is presented in section \ref{subsect:X-ray-radio}.

In this paper we present updated detection statistics and radio photometry of non-thermal sources within the ONC. This continues the non-thermal census presented in Paper 1 of this series by \cite{Forbrich2021} extending their work and investigating the associated radio variability using per-epoch peak flux densities. An accompanying fourth paper will update the astrometry and proper motions continuing the work in Paper 2 of this series by \cite{Dzib2021}, and a fifth paper will investigate intra-epoch variability using these new VLBA epochs continuing the work presented by \cite{Vargas-Gonzalez:2024-thesis}.


\section{Observations and Data Reduction}
\label{sect:obs+data_reduction}

The dataset consists of eight epochs of VLBA observations, each approximately 7 hours in duration. Observations were in the C-band (4-8 GHz) at a central observing frequency of 7.196 GHz in dual polarisation with an aggregate bandwidth of 256 MHz. To minimise the impact of parallax effects on astrometry all observations were scheduled at the same time of the year. The observing setup and data reduction procedures for the first four epochs are described by \cite{Forbrich2021}. The dates of the four new epochs, identified by their project codes BF131A, BF131B, BF131C and BF131D and numbered sequentially as the fifth through eighth epochs are listed in Table \ref{tab:obs_details}. The fifth and sixth epochs, observed in 2020, are separated by four days. While the seventh and eighth epochs, observed in 2021, are separated by one day. This epoch spacing enables the investigation of radio variability on these shorter timescales.

\begin{table*}
	\centering
	\caption{Details of VLBA observations for all eight epochs.}
	\label{tab:obs_details}
	\begin{tabular}{cccclc} 
		\hline
        \hline
       Number & Epoch  & Date, Time (UT)        & Beam$^{1}$      & Antennas & Correlated Positions\\
        \hline
        1 & BF117  & 2015-10-26, 6:25-14:23 & 4.7 $\times$ 1.6 & 8 (no MK, HN) & 556\\
        2 & BF123A & 2017-10-26, 6:23-14:21 & 4.4 $\times$ 1.3 & 9 (no SC, PT$^{2}$, HN$^{2}$) & 557\\
        3 & BF123B & 2017-10-27, 6:19-14:17 & 4.1 $\times$ 1.4 & 8 (no SC, HN, PT$^{2}$) & 557\\
        4 & BF123C & 2018-10-26, 6:24-14:22 & 2.8 $\times$ 1.2 & 10 & 557\\
        5 & BF131A & 2020-10-25, 6:26-14:24 & 2.7 $\times$ 1.1 & 10 & 575\\
        6 & BF131B & 2020-10-29, 6:10-14:08 & 2.7 $\times$ 1.1 & 10 & 575\\
        7 & BF131C & 2021-10-30, 5:37-13:34 & 3.0 $\times$ 1.6 & 9 (no LA) & 575\\
        8 & BF131D & 2021-10-31, 6:03-14:01 & 3.4 $\times$ 1.3 & 9 (no LA) & 575\\
        \hline
    \end{tabular}\\
        \begin{tabular}{l}
        $^1$Synthesised beamsize (FWHM) in milliarcseconds\\
        $^2$Operational but flagged during processing\\
        Columns are (left to right): epoch number, epoch name, date and times of observations, beamsize, antennas, \\
        number of correlated positions.
        \end{tabular}
\end{table*}

The pointing centre of all eight VLBA observations is R.A.=$05^{\text{h}}35^{\text{m}}14.\!^{\text{s}}479$ and Dec.=$- 05^{\circ}22^{\prime}30.\!^{\prime \prime}57$ (J2000), and is the same as that used in the VLA observations of \cite{Forbrich:2016}. As the VLBA and VLA have the same primary beam sizes the VLBA observations effectively cover the same region as the earlier VLA survey.

Phase referencing was performed using the nearby calibrator J0541-0541, located 1.6$^\circ$ in angular distance from the ONC, with a time of 40 seconds on the calibrator and 120 seconds on the target. The calibrator was correlated at the position R.A.$=05^{\text{h}}41^{\text{m}}38.\!^{\text{s}}083384$ Dec. $=-05^{\circ}41^{\prime}49.\!^{\prime \prime}42839$ \footnote{The website (http://astrogeo.org/) provides a catalogue of accurate positions for over 21,000 extragalactic sources with compact radio emission. The positions are updated four times a year, and the current catalogue, rfc$\_$2026b, released on 7 June 2026 lists the position as: \\ R.A.~$=05^{\text{h}}41^{\text{m}}38.\!^{\text{s}}083373 \pm 0.\!^{\text{s}}000006$ Dec.~$=-05^{\circ}41^{\prime}49.\!^{\prime \prime}42861 \pm 0.\!^{\prime \prime}00018$}. This resulted in approximately 4.7 hours of on-source integration time per epoch. Variations in antenna availability between epochs result in the differences in $(u,v)$ coverage and synthesised beam size which are listed in Table \ref{tab:obs_details}.

The data reduction for the newly acquired epochs was carried out using the standard calibration methodology outlined by \cite{Forbrich2021}, employing the Astronomical Image Processing System (AIPS) software suite \citep{Greisen:1988}. As part of the calibration process, systematic delays were mitigated by accounting for ionospheric total electron content, derived from contemporaneous GPS data. Additionally, corrections were applied based on the latest Earth orientation parameters from the US Naval Observatory.

To correct for ionospheric delays three blocks of geodetic observations were obtained per epoch, each 30 minutes in duration. These were observed simultaneously between 4.112 and 7.260 GHz of known calibrator sources from the NRAO calibrator catalogue \footnote{https://obs.vlba.nrao.edu/cst/} spaced across the sky. Residual delays stemming from clock instabilities and zenith atmospheric path differences were subsequently determined through geodetic block observations and subtracted, following the approach of \cite{Reid:2004}. Further calibration included compensating for electronic delays and discrepancies across the IF bands, using the bright calibrator J0530+1331. Finally, phase solutions from the reference source were interpolated and uniformly applied to all individually correlated science targets.

\begin{figure*}
	\includegraphics[width=0.8\paperwidth]{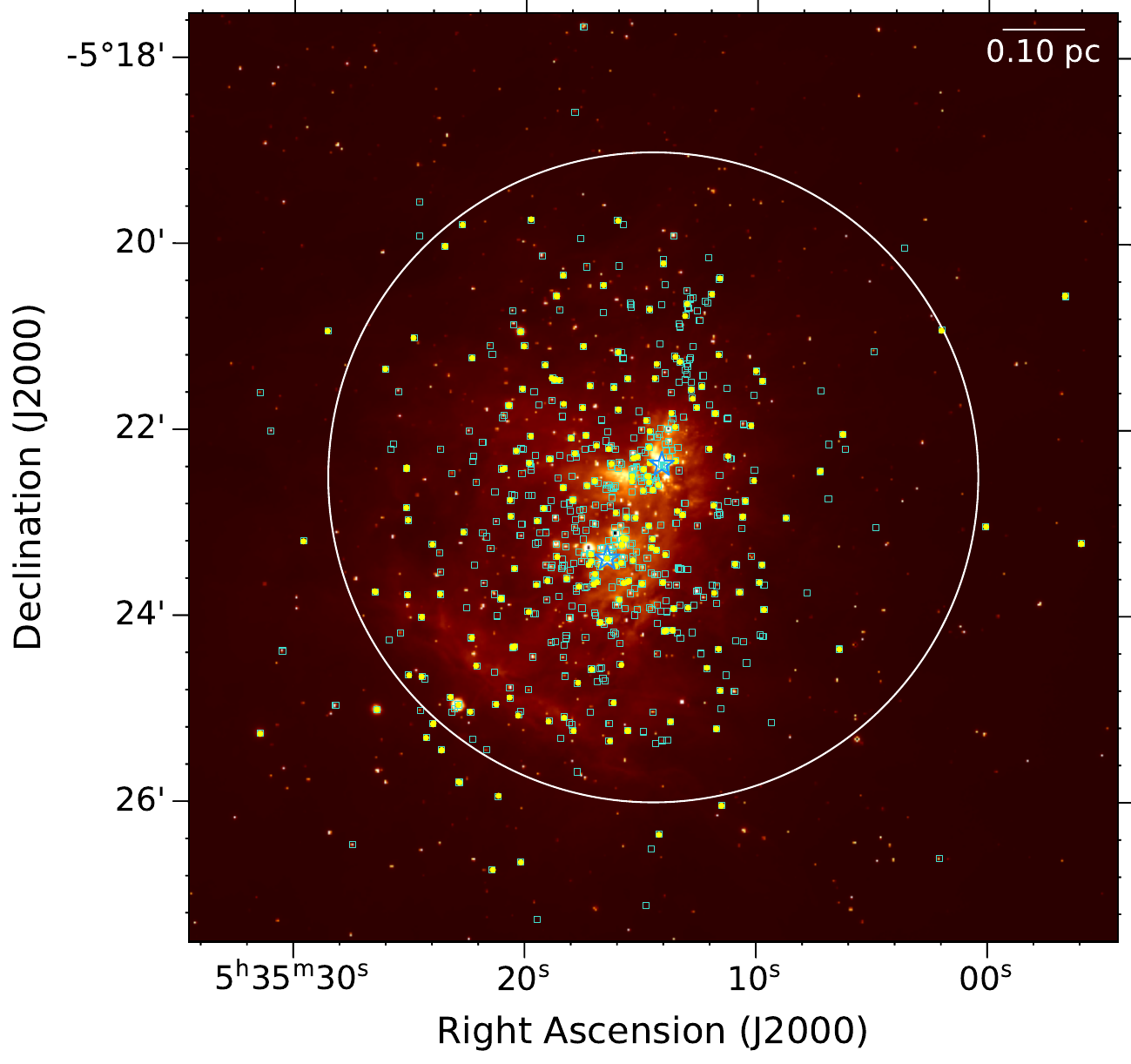}
    \caption{VISION 2.1 $\mu \text{m}$ image of the ONC \citep{Meingast:2016}. The phase centres are marked by the cyan squares, the VLBA source detections by yellow circles and the white circle is the FWHM size of the primary beam ($\sim$7 arcmin). The cyan squares are 4 $\times$ 4 arcsec which corresponds to the larger (8192$\times$8192) image size. The blue stars mark $\theta^1$ Ori C (south) and the Becklin-Neugebauer Object (north).}
    \label{fig:VLBA_correlated_detected}
\end{figure*}

Figure\,\ref{fig:VLBA_correlated_detected} illustrates the spatial distribution of all 575 correlated VLA source positions used in this work overlaid on a VISION 2.1\,$\mu$m image of the ONC. The cyan squares denote the targeted VLBA phase centres and their size corresponds to the initial 4 $\times$ 4 arcsec imaging regions. Radio source detections, defined by a signal-to-noise ratio of $\geq6.5$ from \texttt{imfit} (see Section 3) are marked by yellow circles. The large white circle indicates the FWHM of the VLBA primary beam.


\section{Data Analysis}
\label{sect:data_analysis}

The source detections and statistics for the first four epochs are presented by \cite{Forbrich2021}, and astrometric results are reported by \cite{Dzib2021}. This work extends the analysis to include four additional VLBA epochs from the BF131 project, with a focus on photometry.

All 575 target positions were imaged in each of the four new epochs in Stokes $I$, resulting in 2300 images and a total of 4527 images across all eight epochs. Circular polarisation was investigated by producing Stokes $V$ images for all detections in each epoch, where a detection has a signal-to-noise of $\geq6.5$ as measured by the CASA task \texttt{imfit} \cite[see below for further discussion,][]{CASA2022}.

To identify peaks within a wide field we use images of $8192^2$ pixels with a pixel size of 0.5 mas and natural weighting. This corresponds to an area per image of approximately 16 square arcseconds (corresponding to $\sim\!2.6\times10^6$ square AU at the distance of the ONC) which is sixteen times larger than the image sizes from \cite{Forbrich2021}. The images were centred on the VLA source positions identified by \cite{Forbrich:2016}. This large search area was chosen to maximise the chances of detecting systems with multiple components or sources with significant positional offsets. While most detections lie within tens of milliarcseconds of their VLA position some sources exhibit larger separations. For example, ONC105 (see section \ref{sect:results_discussion} for source nomenclature) was detected at a signal-to-noise ratio (S/N) of 11 at a separation of $673.0 \pm 0.5$ mas, corresponding to approximately 260 AU at the distance of the ONC. These are plausibly associated with components of multiple systems and motivate the decision to not impose an angular separation cutoff during imaging.

The effective field-of-view in VLBI observations is limited by time and frequency smearing and therefore should be considered in the imaging strategy \citep{Wrobel:1995}. Frequency smearing arises because the visibilities measured over a finite bandwidth are gridded as if they were monochromatic, resulting in radial smearing of sources away from the phase centre and a decline in the peak response to a point source. As a consequence both angular resolution and sensitivity degrade with increasing distance from the phase centre. Time smearing also introduces amplitude losses for sources located away from the phase centre. These losses arise from the apparent motion of a source through the coherence pattern due to the Earth's rotation, such that averaging samples in time result in a reduction of measured peak flux density \citep[e.g.,][]{Wrobel:1995, Bridle-schwab:1999, Garrett:1999}.

For the observational setup used here the effective field-of-view is limited to FWHM of $\approx\!5$ arcsec due to time and frequency smearing \citep{Wrobel:1995}. These approximate limits inform the imaging size for this work. The overall imaging methodology described by \cite{Forbrich2021} is adopted in this work. The smallest synthesised beam size across the dataset is $1.1\times2.7$ mas that was sampled with at least two pixels per half-power beamwidth in each axis. At the distance of the ONC this angular resolution corresponds to $\sim$0.4 AU. The coronae of young stars are expected to have sizes of up to tens of solar radii where \cite{Getman:2021} estimate coronal extents in the range 0.5-10 $R_{\odot}$ ($10R_{\odot}\approx0.05$ AU). Therefore, at this resolution the emission is expected to be unresolved.

The choice of signal-to-noise detection threshold was informed by the analyses presented by \cite{Middelberg:2013} and \cite{Herrera:2017}. However, the larger images produced for this work have $\sim$67 million pixels which contain $\sim$4.9 million synthesised beam areas. This substantially larger search space increases the probability of false-positive detections and requires careful consideration of the detection threshold.

The expected number of false-positive detections per image was estimated using the number of independent synthesised beams and the Gaussian tail probability. A single image containing $\sim$4.9 million synthesised beam areas yields an expected number of false positives of $\sim$19 at $5\sigma$, $\sim$0.003 at 6.5$\sigma$, and $\sim$8.6$\times$10$^{-5}$ at 7$\sigma$. When scaled to all eight epochs yields an expected $\sim$0.3 false detections and a false positive rate of $\sim$0.4 sources at 7$\sigma$.

\begin{figure*}
    \centering
    \includegraphics[width=0.87\paperwidth]{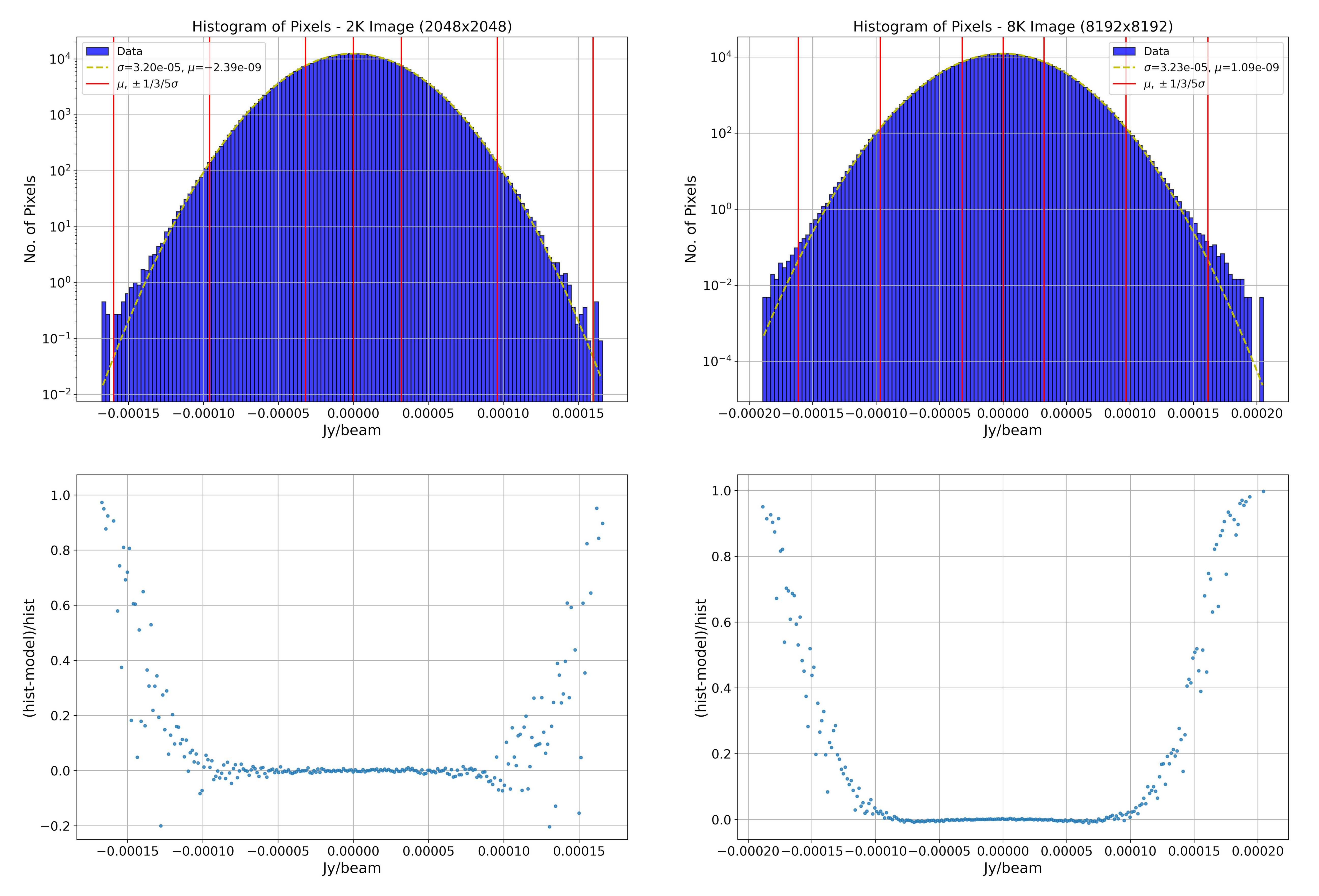}
    \caption{Histogram of the pixels for the 2K images (2048$\times$2048) in left portion and the 8K images (8192$\times$8192) in right portion. Upper panels show the histogram, a model distribution calculated from the pixel mean and standard deviation (yellow dashed line), and red vertical lines indicate the mean and $\pm1\sigma$, $\pm3\sigma$ and $\pm5\sigma$. The bottom panels show the residuals normalised to the histogram. The model shows that the high and low ends exhibit an excess of pixels. We quantify this as an excess kurtosis of 0.02 indicating the bulk is close to Gaussian and tail ratio at 5$\sigma$ of $\sim$4.6 times more extreme pixels showing the significant deviations at the tails.}
    \label{fig:VLBA_histo_4K8K}
\end{figure*}

Figure\,\ref{fig:VLBA_histo_4K8K} presents a histogram of pixel values for both image sizes used in this work with a Gaussian model fitted using the measured pixel mean and standard deviation. The model assumes that the pixels approximately follow a Gaussian distribution. The red vertical lines indicate the mean and $\pm1\sigma,\ \pm3\sigma,\ \pm5\sigma$. The lower panel shows the residuals normalised to the histogram, revealing clear excesses at both low and high pixels values relative to a Gaussian distribution. The consistency between the two image sizes demonstrates that the observed non-Gaussian behaviour is present in the image pixel statistics and is not dependent on the adopted image size. To quantify the deviation we compute the excess kurtosis of 0.02 which indicates that the bulk distribution remains close to Gaussian, although the extreme tails show significant deviations. At the $5\sigma$ level the tail ratio is $\sim$4.6 times more extreme pixels than expected for the nominal Gaussian expectation and at 7$\sigma$ this is effectively 0. This non-Gaussian behaviour implies that detection thresholds of $<5.5\sigma$ are insufficient to identify detections. Consequently, to ensure a negligible false-positive rate in the presence of non-Gaussian noise we adopt a conservative threshold of $7\sigma$.

\cite{Middelberg:2013} also reported a similar non-Gaussian distribution of image noise in their wide-field VLBA observations. They investigated several different possible causes including different visibility weighting schemes and image averaging which did not resolve the behaviour, and were unable to identify an origin. Their observations were conducted at 1.382 GHz and our observations at 7.196 GHz exhibited similar non-Gaussian behaviour, indicating this could be occurring across different observing frequencies. However, we cannot establish whether the effects have the same underlying origin.


    
Peak candidates were initially identified using the peak pixel value and pixel rms measured with the AIPS task \texttt{imstat} \citep{Greisen:1988}. To further motivate the adoption of a high detection threshold we examined the distribution of peak signal-to-noise ratios in the images across the four new epochs. Every image in the fifth epoch contains at least one peak with S/N $\geq$ 5 and at S/N $\geq$ 5.5 a total of 238 peaks were found. The fifth epoch has 28 sources that are real detections (see below for further discussion) meaning that 210 of these are spurious, demonstrating that larger images introduce a substantial number of false-positives.

To ensure real sources were not missed due to elevated noise we also repeated the imaging strategy of \cite{Forbrich2021}. This involved creating smaller undersampled images of $2048^2$ pixels (approximately four million pixels total) which corresponds to between 120,000 and 270,000 synthesised beam areas. Candidate peaks in these images were again identified using \texttt{imstat} adopting a lower threshold of S/N $\geq$ 5.5. These smaller images significantly reduce statistical noise peaks but only allow imaging of a much smaller area of about one square arcsecond. This dual search strategy ensures sensitivity to real sources while maintaining a low false-positive rate.

To summarise the source candidate search strategy two complementary imaging approaches were adopted. First, large undersampled images of $8192^2$ pixels were searched using a conservative threshold of S/N $\geq$ 7. This yielded 189 peak candidates across the four new epochs. Second, to ensure that real detections were not missed due to elevated noise in the larger images, smaller undersampled images of $2048^2$ pixels were also created and searched using a threshold of S/N $\geq$ 5.5 which yielded 122 peak candidates. This follows the methodology of \cite{Forbrich2021}.

Once candidate peaks were identified oversampled images were created for each source candidate in Stokes $I$. These images were $2048^2$ pixels with a pixel size of 0.05 mas, corresponding to an area of 0.1 $\times$ 0.1 arcsec or $\sim$1680 square AU at the distance of the ONC. This image size ensures that each source is fully contained within the imaged area accounting for positional uncertainties in the undersampled images. The pixel scale oversamples the smallest synthesised beam by a factor of 11 ensuring imaging aliasing is negligible. Source positions and flux densities were measured using the CASA task \texttt{imfit} which fits a two dimensional Gaussian \citep{CASA2022}. This task also provides other image statistics such as rms. The data has an average rms sensitivity of 35\,$\mu$Jy\,beam$^{-1}$ and lowest rms of 21\,$\mu$Jy\,beam$^{-1}$ across the four new epochs.

The final detection criteria were defined with the methodology from \cite{Forbrich2021} in mind, where they estimate for the oversampled images which are 2048$^2$ pixels in size that the expected number of false-positive detections is $\sim$0.1 sources across four epochs and the nominal Gaussian expectation is $\sim$0.02. To also maintain consistency with previous work and considering these noise characteristics for these smaller images we also adopt a threshold of S/N $\geq$ 6.5 in the measurements from \texttt{imfit} as our final detection threshold.

Source detection across multiple epochs is also complicated by variability and significant source proper motion. Non-thermal radio emission from YSOs is known to be highly variable \citep[e.g.,][]{Gudel:2002, Deller:2013, Forbrich:2017, Forbrich2021} and the larger images used in this work introduce larger probabilities of statistical noise peaks. In addition, substantial proper motions are present in the ONC.

\cite{Dzib2021} reports proper motions of YSOs within the ONC using the first four epochs of the data used in this work. They report proper motions of up to several milliarcsecond per year meaning sources can shift their position significantly and therefore coincident detections are not possible. As a result detections are assessed on an epoch by epoch basis and associations between epochs are made using source identification from previous VLBA and VLA studies rather than positional coincidence alone.

If epochs are separated by at the most several days then proper motions are negligible and it is assumed that the source is effectively in the same position. This allows epochs that are closely spaced to be concatenated to improve sensitivity. This methodology was adopted by \cite{Forbrich2021} for the second and third epochs which are separated by one day. We extend this strategy by concatenating the fifth and sixth epochs that are separated by four days, and the seventh and eighth epochs separated by one day. In all other respects the imaging strategy for the concatenated data follows that described above.

Once a complete detection list was obtained for all epochs the measured peak flux densities were corrected for primary beam attenuation following \cite{Deane:2024}. The VLBA primary beam is well approximated by an Airy disk with an effective diameter of $D=25.47$\,m \citep{Middelberg:2013}, assumed to be radially symmetric and normalised to unity at the pointing centre. For each source the attenuation factor was calculated from the angular offset at a central frequency of 7.196 GHz and the corresponding correction applied to all flux densities. As discussed by \cite{Deane:2024} uncertainties become increasingly underestimated beyond the $\sim80$~per~cent power point ($\sim$2.8 arcmin), where correction factors grow rapidly, these sources are flagged where appropriate but retained as detections considering their S/N $\geq$ 6.5.

This correction was not applied in the earlier papers in this series of \cite{Forbrich2021, Dzib2021}, whose analyses focused on detection statistics and astrometry. In contrast, our study requires corrected flux densities to derive radio luminosities for the G\"udel-Benz investigation (see section \ref{subsect:X-ray-radio}) and for a per-epoch variability analysis, motivating the application of a primary beam correction.

Finally, all confirmed source detections were also imaged in Stokes $V$ to investigate circular polarisation. These images were produced using the same oversampled imaging strategy described above.

\subsection{Variability Factor}
\label{subsect:VF}

The peak flux densities from \texttt{imfit} across multiple epochs allow an analysis of radio variability on the timescales of the epoch spacing (see Table \ref{tab:obs_details} for epoch dates). We use the variability factor (VF) to quantify source variability, this is defined as the ratio of the maximum to minimum \textit{per-epoch} peak flux densities for a source. Larger values of VF indicate higher levels of variability \citep[e.g.,][]{Vargas-gonz:2023, Vargas-Gonzalez:2024-thesis}.

For sources detected in multiple epochs their VFs were calculated directly from the epoch-averaged peak flux density and uncertainties from standard error propagation of the individual flux density measurements. All the variability measurements in this work are significant with a S/N $\geq$ 3. This ensures that measured variability reflects intrinsic source behaviour rather than fluctuations within the uncertainty from image noise. For sources detected in only a single epoch lower limits of the VF were estimated by adopting $5.5\sigma$ of the lowest rms as an upper limit. This threshold was motivated by peaks of $\leq5\sigma$ that are unreliable due to non-Gaussian noise properties as discussed in section \ref{sect:data_analysis}. Therefore, a more conservative threshold of 5.5$\sigma$ is adopted for upper limit estimates.

In this work we define higher variable sources as those exhibiting changes in their flux density by a factor of $\geq$ 5. This threshold was chosen empirically based on the distribution of inter-epoch VFs observed within the dataset and lies well beyond statistical and calibration uncertainties, this also isolates distinct higher variability observed within the VF distribution.


\section{Results and Discussion}
\label{sect:results_discussion}

Throughout this section we use the term \textit{source} to refer to a unique object and \textit{detection} to denote an individual epoch in which a source is detected, this distinction avoids confusion for sources detected in multiple epochs. From a total of 575 targets we report 226 sources corresponding to 335 detections and an overall detection fraction of 39 per cent. As noted in section \ref{sect:intro} it is not expected that all VLA detected sources could be detected by VLBA observations as some are known to be non-stellar in origin \citep[e.g.,][]{Forbrich:2016, Vargas-gonz:2021}. Most detected sources (188 sources, $\sim$83 per cent of unique sources) are detected in only a single epoch, while 35 sources ($\sim$16 per cent) are detected in multiple epochs and of those 9 sources ($\sim$4 per cent) are detected in all epochs. There are also 19 sources detected in subsequent closely spaced epochs. In addition to the 123 sources reported by \cite{Forbrich2021} we identify 59 sources from the four newly observed epochs and a further 34 sources detected only in the concatenated data, yielding a total of 93 sources. Among the detected sources 177 have S/N $\geq$ 10 and of these there are 46 sources that have S/N $\geq$ 20 and 10 sources that have S/N $\geq$ 50.

It should be highlighted that a significant number of VLA detected sources remain undetected in the VLBA epochs. There are 349 sources ($\sim$61 per cent) detected in the VLA data that are not detected in the VLBA epochs. The average rms sensitivity of the VLBA observations ($\sim\!35\,\mu$Jy beam$^{-1}$) is roughly an order of magnitude higher than for the VLA data ($\sim\!3\,\mu$Jy beam$^{-1}$) reported by \cite{Forbrich:2016}. If we assume the flux densities in the VLA data do not change there are 217 sources that are above the VLBA detection threshold of $S\geq0.2$ mJy beam$^{-1}$ (6.5$\sigma$ of the average rms of the data) and of these there are 134 sources not detected in our VLBA data. However, this difference in source detection cannot solely be explained by this difference in sensitivity. Instead, this could represent a subset of VLA detected sources lacking a compact non-thermal component, variability, the difference in sensitivity discussed above or a combination thereof. It should be highlighted that VLBA non-detections cannot be assumed to be thermal sources, as strong variability within the VLBA population means that some non-thermal sources may lie below the VLBA detection threshold during the epochs of our data. The source populations are also discussed by \cite{Forbrich:2016} from their VLA observations.

The source identifiers are from the [FRM2016] catalogue presented by \cite{Forbrich:2016} of 556 sources and are labelled as `ONC' followed by their catalogue number. The 19 additional sources reported by \cite{Vargas-gonz:2021} are appended sequentially to this catalogue, for example the first additional source is designated as `ONC557'. We consider a source associated with the nominal VLA position when the angular separation is $\leq$ 800 mas. This threshold is motivated by the range of separations observed in our sample. For example, ONC105 is detected at S/N = 10.9 at a separation of 673.0~$\pm$~0.5~mas (corresponding to $\sim$261 AU at the distance of the ONC).

There are detection candidates that are at larger separations of $>$800 mas, for example, there is a candidate for ONC229 within the sixth epoch at S/N = 19.5 at a separation of 1025.7~$\pm$~0.1~mas ($\sim\!398$ AU). This candidate has the highest signal-to-noise of these potential candidates. Sources at such large separations from their respective phase centre could indicate a binary nature to these systems or coincident alignments of unrelated sources. These detection candidates found at larger separations are discussed in Appendix \ref{appxA}.

We also identify cases where multiple VLBA sources associated with the same phase centre are detected. In total 26 such sources are identified and listed in Tables \ref{tab:flux_table} and \ref{tab:source_detections}. Sources are classified as separate when their angular separations are too large to be explained by plausible proper motions or when multiple components are detected simultaneously within the same image. In these cases they are numbered sequentially based on the order they are detected (e.g. ONC002.1 is detected first within the third epoch and ONC002.2 is detected second within the fourth epoch). Where sources are detected simultaneously that have not been detected previously the brightest source (i.e. the highest peak flux density) is designated as the `first' source. For example, ONC427 is detected simultaneously as two separate sources and neither have previously been detected therefore the source with the higher flux density is designated as ONC427.1 and the other source as ONC427.2.

Four source pairs are detected within the same image: ONC177 in epoch 4 at a separation of $22.9 \pm 0.3$ mas ($\sim\!9$ AU), ONC413 in the concatenated epoch of BF131AB at a separation of $16.6 \pm 0.4$ mas ($\sim\!6$ AU), ONC414 in epoch 4 at a separation of $168.2 \pm 0.3$ mas ($\sim65$ AU) and ONC427 in epoch 7 at a separation of $16.6 \pm 0.4$~mas ($\sim\!6$ AU). We note it is currently unknown whether these sources are gravitationally bound. These sources will be discussed in more detail in a future astrometry paper.

\subsection{Circular Polarisation}
\label{subsect:polarisation}

Stokes $V$ images were created for all detections in each epoch to search for circular polarisation. Previous work has only occasionally detected circular polarisation, for example \cite{Zapata:2004-1} detects varying circular polarisation from 4 to 10 per cent associated with one of the 77 sources (associated with ONC066 or GMR A) detected across their four VLA observing epochs at 3.6 cm. \cite{Zapata:2004-2} also detected approximately 20 per cent circular polarisation from one of the 11 sources (associated with ONC154 or their source 140-410) in their VLA observations at 1.3 cm wavelength. These measurements provide useful comparisons although the VLA and VLBA have different spatial sensitivities and probe different angular scales, in addition to differences in observing frequency and sensitivity.

No significant Stokes $V$ detections were identified within our VLBA sample. Therefore, upper limits on the fractional circular polarisation were estimated using the rms of the Stokes $V$ images, following $\text{n}\ \sigma_V / I_{\text{peak}}$, where $n=5.5$, $\sigma_V$ is the Stokes $V$ rms and $I_{\text{peak}}$ is the Stokes $I$ peak flux density. The resulting upper limits span a wide range from 1 up to 100 per cent with a median upper limit of $\sim\!13$ per cent. Upper limits approaching 100 per cent do not imply highly circular polarised emission, rather they indicate the source is too faint relative to the Stokes $V$ noise for a meaningful constraint. The best constrained upper limit is associated with the brightest detection in our data of ONC254 in the first epoch with an upper limit of $\sim\!1$ per cent. For ONC066 our $5.5 \sigma$ upper limits on the fractional circular polarisation range from 2 to 12 per cent and therefore do not exclude the 4 to 10 per cent circular polarisation previously detected by \cite{Zapata:2004-1}. Similarly, our upper limits for ONC154 range from 7 to 73 per cent, compatible with the $\sim20$ per cent circular polarisation reported by \cite{Zapata:2004-2}.

Therefore, the absence of Stokes V detections does not rule out gyrosynchrotron emission or circular polarisation and for sources with better constrained upper limits still provides a constraint on the degree of polarisation. It is also worth highlighting that circular polarisation can be highly variable and would only represent a fractional percentage of the total intensity \cite[e.g.,][]{Dulk:1985, Feigelson:1998, Kaur:2024}.

\subsection{Spectral Types and Evolutionary Classes}
\label{subsect:sp type ev classes}

We searched for stellar spectral classifications associated with all detections by cross-matching with the catalogue of \cite{Hillenbrand:2013}, which contains spectral classifications for approximately 600 sources in the ONC. A search radius of 0.5 arcsec was adopted to account for significant proper motions and potential positional offsets between the radio sources and counterparts at other wavelengths. We note that positional coincidence does not necessarily imply physical association particularly in the case of multiple systems. Using these criteria 64 of our detected sources have reported spectral classifications.

Spectral classifications are available for only $\sim$28 per cent of the VLBA source population and therefore we do not attempt to draw conclusions about the spectral-type distribution of the full sample. Within the classified subsample, sources span a wide range of spectral types including M, K and K-M types. The highly variable sources with spectral classifications are likewise M, K and K-M types with two notable exceptions. ONC254 (spectral type O-B) which is a known multiple system \citep{Lohsen:1976, Bossi:1989, Petr:1998} where the non-thermal radio emission is associated with a T Tauri companion \citep{Schertl:2003}. ONC480 (spectral type B) which is discussed by \cite{Dzib:2026b} as an astrometric binary and the non-thermal radio emission may similarly originate from a lower mass companion.

We also searched for protostellar evolutionary classes by cross-matching with the X-ray and near-infrared study of the inner ONC by \cite{Prisinzano:2008} which provides evolutionary classes for X-ray detected YSOs. Using the same search parameters as for the spectral classifications we find 4 of our sources are classified as class II and 8 sources as class III. However, as noted by \cite{Forbrich2021}, fewer than 5 per cent of the VLA sources from \cite{Forbrich:2016} have counterparts in \cite{Prisinzano:2008} and consequently, meaningful detection fractions for evolutionary classes cannot be derived. It should be noted that in section \ref{subsect:highest_VFs} and in Appendix \ref{appxB} we additionally include details from other references in the discussion of higher variable sources where appropriate.

A powerful method for identifying YSOs is through association with X-ray sources using the COUP X-ray survey as a reference, as outlined in section \ref{sect:intro}. We refer to \cite{Forbrich:2016, Forbrich2021, Vargas-gonz:2021} who searched for COUP associations of the radio source positions used in this work, who also use a search radius of 0.5 arcsec. Of the 226 sources detected in this work 126 have COUP counterparts ($\sim$56 per cent), and of these 36 of the VLBA radio sources are newly detected in this work identifying these YSOs as having a detectable non-thermal radio component. 100 VLBA radio sources ($\sim$44 per cent) do not have COUP counterparts and of these 37 are newly detected VLBA sources.

Additionally, to identify sources as being associated with the ONC we cross matched our source list with the \textit{Hubble} Source Catalog \citep{Whitmore:2016} a master catalogue from archival HST imaging with the ACS, WFPC2 and WFC3 instruments. The VISTA near-infrared survey of the Orion A molecular cloud, called VISION, covering $\sim$18.3 square degrees reported by \cite{Meingast:2016} was also cross matched. Here we also used a search radius of 0.5 arcsec for both catalogues. There are 138 VLBA sources that have \textit{Hubble} counterparts and 98 sources that have VISION counterparts, of these there are 40 sources that have \textit{Hubble} and/or VISION counterparts that do not have COUP counterparts. For example, ONC476 (V* V2370 Ori) does not have a COUP counterpart but is coincident with a \textit{Hubble} and VISION source within $\sim$0.2 arcsec and is classified by \cite{Fang:2021} as an M5 star. There are 37 sources that lack any counterpart in the other catalogues (i.e. these sources are only detected in the radio). As a final check we compared source positions to a JWST image (NIRCAM general purpose F277W filter)\footnote{https://jwst.esac.esa.int/archive/} and found one association between our VLBA sources that lack any counterparts and a JWST source which is ONC551 at a separation of $\sim\!0.2$ arcsec.

We additionally cross matched our VLBA source positions with the \textit{Hubble} treasury program catalogue using the ACS/WFC of proplyds and circumstellar structures presented by \cite{Ricci:2008}, here we again use a search radius of 0.5 arcsec. We find 28 associations between our VLBA sources and sources within their catalogue, and find that 19 of these have COUP counterparts and 9 sources do not. These associations demonstrate that compact nonthermal radio emission is present in a number of systems associated with irradiated circumstellar structures in the ONC. It is important to highlight that the emission detected by the VLBA is distinct from the extended thermal emission associated with the ionised proplyd, this indicates a compact nonthermal component associated with the underlying YSO.

To assess whether these remaining sources lacking counterparts could be background sources we estimate the expected number of extragalactic sources based on the work by \cite{Windhorst:1993}, following the same methodology as \cite{Forbrich2021}. We estimate up to 3 extragalactic radio sources at $S\geq0.2$ mJy beam$^{-1}$ (or 6.5$\sigma$ of the average rms of the data) within the primary beam. However, if we instead consider the area that has been imaged in this work (i.e. the larger image sizes and the 575 phase centres which corresponds to 9200 square arcsec) we estimate $\sim$0.2 background sources. It is also worth highlighting that this is based on VLA data and as VLBI observations are only sensitive to nonthermal emission this estimate represents an upper limit. Therefore, background contamination is considered negligible and most of these sources are likely members of the ONC. This is also consistent with the estimates and conclusion of \cite{Forbrich2021}.

\subsection{Radio Variability}
\label{subsect:radio var}

Variability was quantified using the variability factor (VF) as defined in section \ref{subsect:VF}, derived from \textit{per-epoch} peak flux densities. We identify 21 higher variable sources ($\sim$9 per cent of unique sources) where 11 sources have measured VFs $\geq$ 5 and a further 10 sources have lower limits $\geq$ 5. We also highlight that 4 sources ($\sim$2 per cent) exhibit extreme variability with VF $\geq$ 15, these are discussed below in section \ref{subsect:highest_VFs}. Table \ref{tab:flux_table} lists the flux densities, COUP numbers and inter-epoch VFs, and Table \ref{tab:source_detections} lists additional source information: signal-to-noise, source positions, separation from nominal VLA position, YSO class and spectral type. The sources exhibiting higher VFs all have COUP counterparts identifying them as YSOs and some have YSO evolutionary classes, see section \ref{subsect:highest_VFs} and appendix \ref{appxB}.

\begin{figure*}
	\includegraphics[width=0.8\paperwidth]{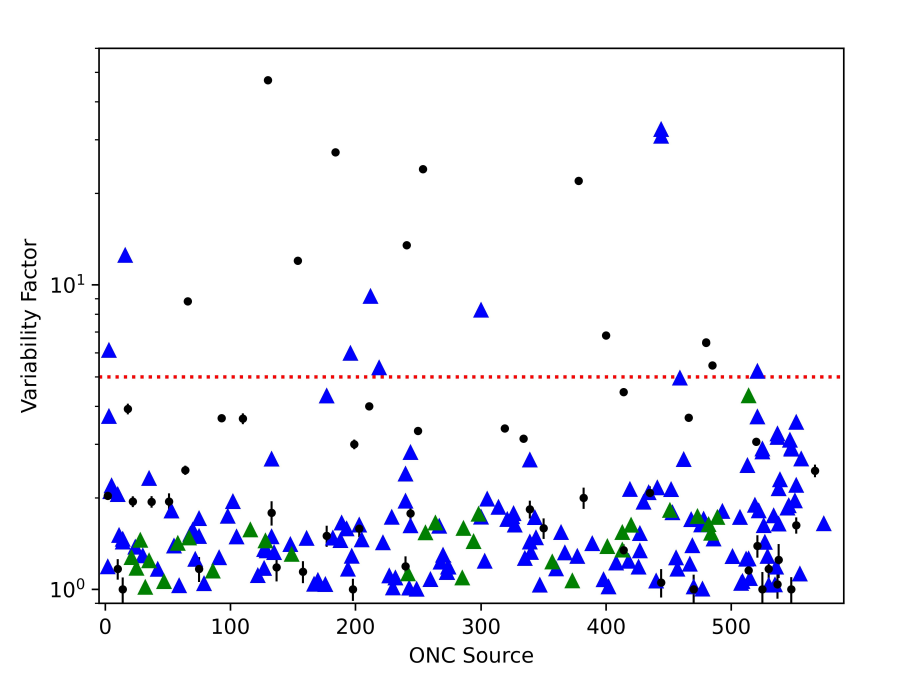}
    \caption{A plot of all variability factors from the eight epochs of VLBA data. Variability factors represent the ratio between the highest and lowest flux densities where higher numbers indicate increased variability. Sources that are detected in multiple epochs enable the variability factor to be calculated directly (black points). Whereas, sources that are only detected in a single epoch a lower limit for the variability factor was calculated using the 5.5$\sigma$ value of the lowest rms noise as an upper limit (blue triangles for single detections in the normal data and green for the concatenated data). The red dotted line marks the limit for higher variability of VF $\geq$ 5, see section \ref{subsect:VF}.}
    \label{fig:VLBA_VF_plot}
\end{figure*}

Figure\,\ref{fig:VLBA_VF_plot} displays the maximum VF for sources detected more than once (black points), lower limits for sources detected in only a single epoch (blue triangles) and lower limits for sources detected solely in the concatenated data (green triangles). The red dotted line indicates the threshold of higher variability of VF = 5. Sources lying on this threshold are classified as highly variable if their uncertainty overlaps VF = 5 or when they are constrained by lower limits. For example, ONC414.1 has a VF = 4.5 $\pm$ 0.6 and is therefore consistent with VF $\geq$ 5 within uncertainties. We note, as discussed in section \ref{subsect:VF}, that all variability factors derived in this work are significant with S/N $\geq$ 3. Additionally, we examined distributions of VFs for the 54 total sources with directly derived VFs to search for evidence of breaks or bimodal structure. No clear evidence for either is apparent, although the relatively small number of sources limits the sensitivity of this test.

Several sources previously reported as radio-variable are detected in our data but do not exhibit high variability over the epochs sampled here. These include the `ORBS' source (ONC198) reported by \cite{Forbrich:2008} and sources ONC053 (COUP 427), ONC110 (COUP 530), ONC469 (COUP 1101) and $\theta^2$ Ori A (ONC515 or COUP 1232) reported by \cite{Forbrich:2017}. \cite{Vargas-Gonzalez:2024-thesis} discusses sources ONC022, 326, 435, 547 and 557 being detected at higher signal-to-noise within their VLA data. Of these ONC022 and 557 have VFs $\geq$ 5 in their work and were not classified as variable by \cite{Forbrich:2017}.

These sources being detected in our VLBA data enable their VFs to be evaluated within the framework of this study. All of these sources mentioned above have VFs $<$ 5. These variability factors when compared to earlier reports of high variability underscore the importance of multi-epoch observations for characterising intermittent radio emission as well as the need for closely spaced epochs to probe shorter timescales.

Additionally, we report the second VLBI detection of the variable source $\theta^1$ Ori C (ONC305) in the eighth epoch at a peak flux density of $0.24 \pm 0.02$ mJy beam$^{-1}$ (S/N = 12.3). This source is only detected in one out of the eight epochs analysed here indicating an intermittent source. The first VLBI detection of this source is reported by \cite{Dzib:2026a}, where they report the radio detection at a separation of 14 mas ($\sim$5 AU) from the optical source. This source has a known companion with an orbital period of 11.26 $\pm$ 0.05 yr and semimajor axis of 44 $\pm$ 3 mas reported by \cite{Kraus:2009}. The companion has a mass estimate of $\sim$11 M$_{\odot}$ \citep{Balega:2014} which is well above the expected mass range for magnetically active young stars. \cite{Dzib:2026a} discusses that considering the orbital configuration reported by \cite{Kraus:2009} that the radio emission is likely associated with the companion. However, additional VLBI observations are required to study and confirm the nature of this source.

We also report a single VLBI detection of source ONC441 in the seventh epoch at a flux density of $0.29 \pm 0.02$ mJy beam$^{-1}$ (S/N = 12.2), with a lower limit of the VF of $>2.2$. This source was reported as a radio source by \cite{Forbrich:2016} and has previously been identified as a radio counterpart to a component of a proposed Jupiter Mass Binary Object (JuMBO) by \cite{Rodriguez:2024}. These authors find a persistent VLA counterpart across multiple observations. Our detection provides evidence that the source also produces nonthermal radio emission. While the nature of this source remains unclear, near-infrared analysis by \cite{Luhman:2024} indicates that this source is a close pair of low-mass stars.


\subsection{Highest Variability Factors}
\label{subsect:highest_VFs}

In this section we discuss the sources exhibiting the most extreme variability factors, here we highlight sources with VFs $\geq$ 15 to illustrate the highest VFs in this study which totals four sources. Additional highly variable sources are described in Appendix \ref{appxB}. Figure \ref{fig:VF5_img_grid} shows continuum images for the four sources discussed below showing their maximum and minimum flux densities.

\subsubsection{ONC130 (COUP 554 or 2MASS J05351356-0523552)}
ONC130 is classified as a YSO with a disk by \cite{Megeath:2012} from their Spitzer mid-infrared observations, and as a class I YSO by \cite{Grosschedl:2019} using near-infrared data from the ESO-VISTA Orion A survey (VISION). Figure\,\ref{fig:higher_VF_LCs} shows the light curve of ONC130 (upper left panel), presenting the per-epoch peak flux densities and upper limits (red arrows) for epochs with non-detections. The light curve shows a significant increase in the flux density in the first four epochs peaking in the fourth which are separated by various timescales as seen by the epoch dates at the bottom of the light curve. This source is detected in a total of five epochs with a highest flux density of 8.20 $\pm$ 0.22 mJy beam$^{-1}$ and a lowest in the fifth epoch of 0.18 $\pm$ 0.02 mJy beam$^{-1}$. This corresponds to VF = 47.1 $\pm$ 6.4 which is the highest observed in our data.

\subsubsection{ONC184 (COUP 639 or ALMA J053514.5010-052238.674)}
ONC184 is classified as a variable YSO based on observations from the Gould's Belt VLA Survey Program \citep{Dzib:2013}, and \cite{Kounkel:2014} report a variability percentage of 60.7$\pm$16.1. Additionally, \cite{Getman:2005} and \cite{Rivilla:2015} discuss this source as a flaring object in X-ray and radio respectively. The light curve for this source is shown in the upper right panel of Fig.~\ref{fig:higher_VF_LCs} and exhibits a pronounced increase in flux density during the fifth epoch. ONC184 is detected in all VLBA epochs with a highest epoch-average peak flux density of 5.19 $\pm$ 0.08 mJy beam$^{-1}$ in the fifth epoch, and a lowest peak flux density of 0.19 $\pm$ 0.02 mJy beam$^{-1}$ in the first epoch. This corresponds to VF = 27.3 $\pm$ 2.9.

\subsubsection{ONC254 (COUP 745 or $\theta^1$Ori A$_2$)}
ONC254 is a member of the well-studied multiple system $\theta^1$Ori~A. The primary component ($\theta^1$Ori A$_1$) has a spectral type B0.5 as reported by \cite{Levato:1976} from their Kitt Peak 0.9-m Cassegrain spectrograph observations. It has mass estimates of 18.91 $M_\odot$ by \cite{Hillenbrand:1997}, 20 $M_\odot$ from the SAO 6 m telescope data reported by \cite{Weigelt:1999}, and $14 \pm 5 \text{ M}_{\odot}$ by \cite{Simon-Diaz:2006} from their Isaac Newton 2.5 m Telescope (INT) data. \cite{Lohsen:1976} reports an eclipsing binary with a period of 65.43 days from Bochum 0.61 m telescope observations, and \cite{Abt:1991} derived a period of 65.09 $\pm$ 0.07 days using the 2.1 m Cassegrain spectrograph. This tight companion is $\theta^1$Ori A$_3$ and \cite{Bossi:1989} reports from the thermal spectrum that this companion is at a separation of 0.71 AU and is a T-Tauri star with a mass of 2.5-2.7 M$_\odot$.

\cite{Petr:1998} identifies another companion ($\theta^1$Ori A$_2$)  at a separation of $\sim$200 mas (corresponding to $\sim$80 AU at the distance of the ONC), and they speculate that the radio emission is likely from this companion. This was confirmed by \citet{Dzib2021} through a comparison of VLBA and \textit{Gaia} astrometry for associated sources. The mass of $\theta^1$Ori A$_2$ is reported by \cite{Schertl:2003} as 4 M$_\odot$ with a period of 214 yr orbiting the tight binary $\theta^1$Ori A$_1$+$\theta^1$Ori A$_3$, from their SAO 6 m telescope observations. \cite{gravity:2018} confirmed this third companion is gravitationally bound to the system from their observations using the GRAVITY instrument at the Very Large Telescope Interferometer.

$\theta^1$Ori A$_2$ is detected in all eight epochs and exhibits strong variability. The light curve of this source is shown in the lower left panel of Fig. \ref{fig:higher_VF_LCs} and shows considerable variability across epochs. The highest epoch averaged peak flux density is 42.62 $\pm$ 0.87 mJy beam$^{-1}$ in the first epoch, and lowest of 1.77 $\pm$ 0.04 mJy beam$^{-1}$ in the third epoch. This corresponds to VF = 24.0 $\pm$ 0.7. Interestingly, this source exhibits lower but still high variability of VF = 5.2 $\pm$ 0.1 on a time-scale of just 1 day. The extreme variability observed across different timescales suggests variability on shorter timescales but is difficult to characterise due to the complexity of this source.

\subsubsection{ONC378 (COUP 932 or MT Ori)}

ONC378 is detected in all epochs. It is classified by \cite{Hillenbrand:2013} as spectral type G-K3. Mass estimates for this source vary across the literature: 0.387 $\pm$ 0.022 $M_\odot$ by \cite{Wei:2024} using the `MIST' stellar evolutionary model from their near-infrared `NIRSPEC' work from Keck II 10 m telescope observations \citep{Choi:2016, Dotter:2016}, whereas \cite[][see their Appendix C]{Getman_Feig:2021} gives a mass of 3.7 $M_\odot$ from their \textit{Chandra} data, this is also discussed by \cite{Getman:2022}.

\begin{figure*}
    \centering
    \includegraphics[width=0.65\paperwidth]{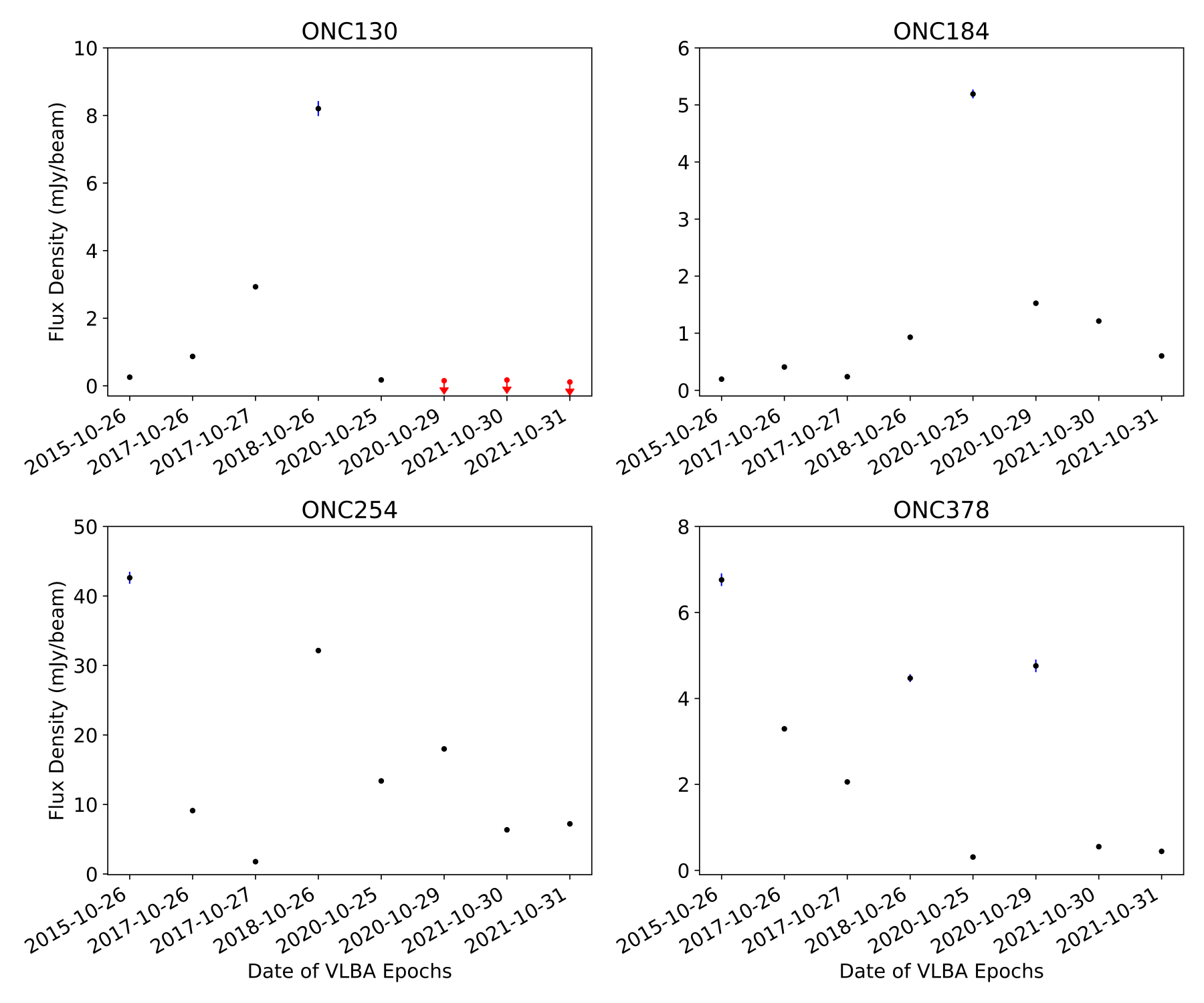}
    \caption{Light curves for the four sources with the highest VFs (VFs for each source listed in the upper right of each panel) using the peak per-epoch flux density. The dates of each epoch are shown and per-epoch peak flux density. Where a source is not detected in an epoch we show upper limits as red arrows.}
    \label{fig:higher_VF_LCs}
\end{figure*}

The light curve for this source is shown in the lower right panel of Fig. \ref{fig:higher_VF_LCs} and displays considerable variability in the per-epoch peak flux densities. The highest peak flux density is 6.76 $\pm$ 0.15 mJy beam$^{-1}$ in the first epoch while the lowest of 0.31 $\pm$ 0.02 mJy beam$^{-1}$ is in the fifth epoch. This corresponds to VF = 22.0 $\pm$ 1.5. Interestingly, this source exhibits high variability on a shorter timescale of 4 days of VF = 15.3~$\pm$~1.1. As discussed in section \ref{subsect:VFs_across_time-scales} this source exhibits a range of variability across different timescales with markedly different extremes measured for the same object.


\begin{figure*}
    \centering
    \includegraphics[width=0.55\paperwidth]{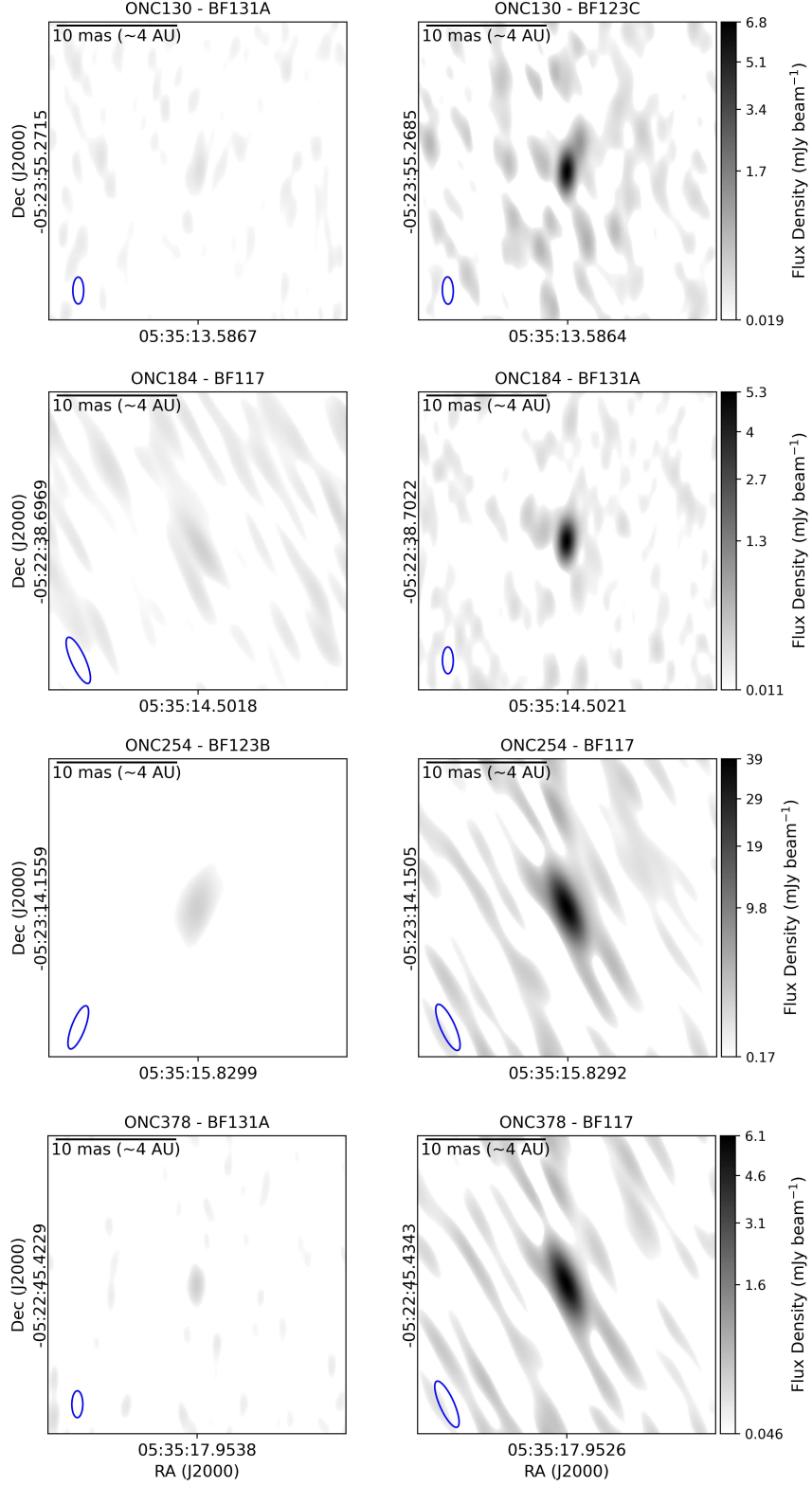}
    \caption{Grid plot of radio continuum images showing the four sources with VF $\geq$ 15. The RA and Dec of the centre of the image are shown, in the bottom left of each image is the synthesised beam and upper left is a scale bar (10 mas corresponds to $\sim$4 AU at the distance of the ONC). The images are 512x512 in size with a pixel size of 0.05 mas to highlight the source. A colourbar is to the right of each row showing flux density in mJy beam$^{-1}$ illustrating the range associated with these highly variable source detections.}
    \label{fig:VF5_img_grid}
\end{figure*}


\begin{figure*}
	\includegraphics[width=0.7\paperwidth]{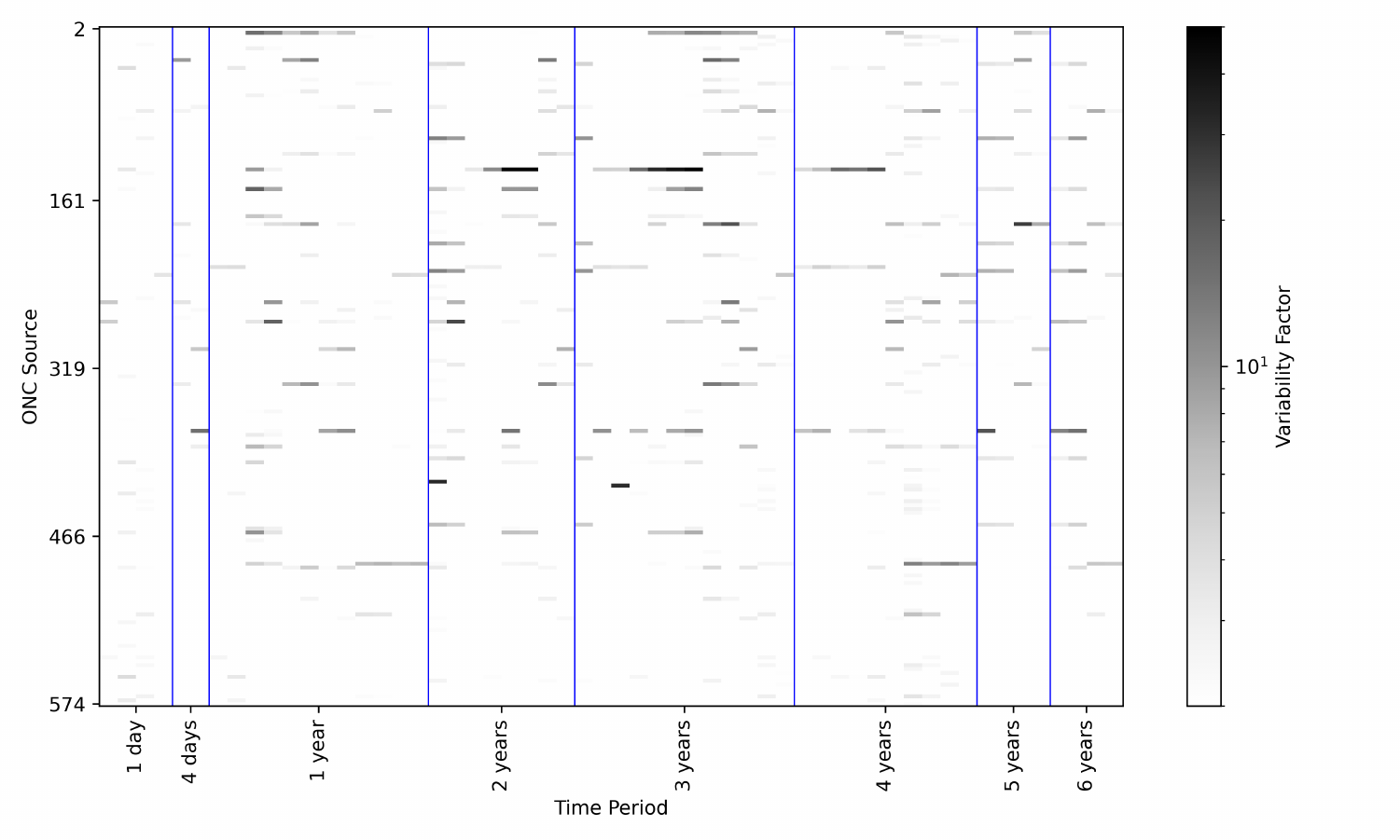}
    \caption{A heatmap showing the VFs $\geq2$ per source that are separated into time-scales. The y axis indicates the source identifier. The columns are sorted into the respective time periods between epochs, and the vertical blue lines indicate the boundaries of each time period. The highest derived VF is associated with ONC130 at VF = 47.1 $\pm$ 6.4.}
    \label{fig:VLBA_ALL_VFs_heatmap}
\end{figure*}

\subsection{Source and Variability Occurrence}
\label{subsect:VFs_across_time-scales}

Multi-epoch observations enable the characterisation of radio variability across a wide range of timescales, and the intermittent nature of non-thermal radio emission makes such an approach essential. By using per-epoch peak flux densities we probe variability on timescales ranging from several years to one day, see Table \ref{tab:obs_details} for epoch dates.

Figure\,\ref{fig:VLBA_ALL_VFs_heatmap} shows the variability factors that are higher than 2 for all sources at all available time separations between epochs as a heatmap, this highlights not only the highest variability but also where source's fluxes vary by at least a factor of 2. The 11 most variable sources (VFs $\geq$ 5) that have multiple detections exhibit a variety of VFs across the different timescales between epochs. For example, ONC130 has a maximum VF of 47.1 $\pm$ 6.4 on a timescale of two years, compared to 9.4 $\pm$ 0.3 on a one year timescale and 3.4~$\pm$~0.1 on one day. Similarly, ONC241 exhibits a highest VF of 13.8~$\pm$~1.4 on a timescale of three years, while the second highest VF is 9.8 $\pm$ 1.0 on a timescale of just 1 day.

The presence of such a broad range of variability factors for individual sources across differing timescales, as illustrated in Fig. \ref{fig:VLBA_ALL_VFs_heatmap}, indicates a substantial fraction of the radio variability likely occurs on timescales shorter than those sampled in this work. This would also be consistent with previous studies of flaring associated with YSOs at higher time resolutions. For example, \cite{Bower:2003} reports a bright YSO millimetre flare (GMR-A or ONC066) in the ONC on a timescale of several hours and \cite{Forbrich:2008} reports the first bright flare observed at centimetre wavelengths (the `ORBS' source or ONC198) on a timescale of several hours. \cite{Forbrich:2017} then reports several flares occurring on timescales from hours down to $\sim$30 minutes. Additionally, \cite{Vargas-Gonzalez:2024-thesis} reports variability at millimetre wavelengths for several sources and for two sources (ONC098 and 130) from the first four epochs of the VLBA data used in this work, both on timescales of hours.

We now investigate the detectability and variability of YSOs across the eight epochs of this work with the aim of quantifying how often these sources are detected and how the radio emission varies between epochs. We consider the 254 sources with COUP counterparts that lie within the adopted search radius of their respective phase centre to take a conservative approach that ensures these are YSOs (see discussions in sections \ref{sect:intro} and \ref{sect:results_discussion}). There is additionally evidence that sources without X-ray counterparts are associated with YSOs and a discussion of all sources in the context of their identification as ONC members is in section \ref{subsect:sp type ev classes}, therefore, the approach taken here results in a lower limit. Of the 254 sources considered here 126 are detected within our VLBA data totalling 250 detections across all eight epochs.

The probability of detecting a YSO in a given epoch can be estimated from the total number of detections divided by the total number of sources with COUP counterparts and the number of epochs, $\rho = 250 / (254 \times 8) \approx12$ per cent. This indicates that $\sim\!12$ per cent are detected in any individual epoch. In addition, there are 92 sources that are detected only once. Considering these together highlights that YSO radio emission is highly intermittent.

To assess how long this emission persists we consider detections in adjacent epochs, where adjacent epochs means all subsequent epochs (e.g. the first and second epoch are adjacent and separated by 2 years, as seen in Table \ref{tab:obs_details}). There are 25 sources detected in adjacent epochs and of these 19 sources are detected in closely spaced epochs (closely spaced epochs are separated by 1 or 4 days) which corresponds to a probability of $\sim\!9$ per cent. The relative rarity of detections across closely spaced epochs suggests that higher variability is occurring on timescales shorter than the shortest spacing between epochs.

We now consider the occurrence of higher variability defined as VF $\geq5$, where we find 21 sources which corresponds to $\sim$8 per cent of VLBA detected sources. Of these, 12 sources exhibit VF $\geq5$ between adjacent epochs, indicating that $\sim$5 per cent of the source sample shows high variability between adjacent epochs, this also corresponds to $\sim$57 per cent of the subsample of higher variable sources.

Finally, we investigate closely spaced epochs (separated by 1 or 4 days). Of the 21 sources that are considered highly variable there are 3 sources (ONC241, 254 and 378 which are all associated with YSO sources, see section \ref{subsect:highest_VFs} and Appendix \ref{appxB}) that have VFs $\geq5$ between these closely spaced epochs ($\sim\!14$ per cent of the subsample of highly variable sources). These cases demonstrate that large amplitude variability can occur over the shortest epoch spacing of 1 day.

Considering these results together show that YSO radio emission is both highly intermittent and strongly variable. The low probability of detecting a source in a given epoch combined with the low likelihood of detections in closely spaced epochs further suggest that YSO variability is likely occurring on timescales shorter than our shortest epoch spacing of 1 day.

\subsection{Radio - X-ray Connection}
\label{subsect:X-ray-radio}

The non-thermal gyrosynchrotron radio emission in YSOs originates from mildly relativistic electrons associated with coronal-type activity. Thermal bremsstrahlung X-ray emission occurs from heated plasma within the corona \citep{Dulk:1985, Feigelson:1999, Gudel:2002, Forbrich:2013}. There is an empirical relation between non-thermal radio and thermal X-ray emission first studied for coronally active stars by \cite{Gudel-benz:1993} and \cite{Benz-gudel:1994}, the `G\"udel-Benz (G-B) relation'.

This relation has previously been studied for the ONC using VLA and COUP data by \cite{Forbrich:2013}, and by \cite{Yanza:2022} who also used VLA and COUP data but additionally used the VLBA source detections from \cite{Forbrich2021} as a selection criterion of their VLA sources. These studies were somewhat limited as it is difficult to disentangle thermal and non-thermal emission within VLA observations (see section \ref{sect:intro}). However, these studies both found that their data are compatible with the relation. We therefore investigate this using VLBA data that provides the unique perspective of being sensitive to exclusively non-thermal emission.

Here we use the COUP X-ray luminosities as the COUP catalogue is the most sensitive X-ray survey of the ONC and this high sensitivity allows the spectral fitting required to derive total band (0.5-8 keV) absorption corrected luminosities which are corrected for interstellar absorption as discussed by \cite{Getman:2005}. We then derived radio luminosities from our multi-epoch VLBA data. We highlight that these datasets are not simultaneous and therefore variability in one dataset would not be observed in the other, this methodology of using non-simultaneous data is also extensively discussed by \cite{Forbrich:2013}. The empirical correlation reported by \cite{Gudel-benz:1993} and \cite{Benz-gudel:1994} is:

\begin{equation}
    \frac{L_{X}}{L_{R}} = k\cdot10^{15.5\pm0.5}
    \label{eq:gudel-benz}
\end{equation}

\noindent where $k\leq1$ accounts for different types of stars. \cite{Benz-gudel:1994} give $k \approx 0.17$ for RS CVn binaries, Algol systems, FK Com stars and Post T Tauri stars. Whereas, for YSOs other studies have shown that $k$ is significantly less than 1. For example, \cite{Dzib:2015} gives $k = 0.03$ for YSOs of several star-forming regions.

There are 126 of our VLBA sources that have COUP counterparts totalling 250 detections, so we use the luminosities for these sources. We also correct the X-ray luminosities to the distance used in this work as \cite{Getman:2005} used an estimate of 450 pc. We then use a linear least-squares fit in log-log space, corresponding to a power law scaling between radio and X-ray luminosities, to our 250 total detections which gives $\log{L_X} = (12.5 \pm 1.6) + (1.08 \pm 0.10) \log{L_R}$ with a Pearson correlation coefficient (R) of 0.58. The Pearson coefficient of 0.58 indicates a moderate positive correlation, which is statistically significant (p $<\!0.001$). Despite the significant association, the moderate magnitude of the correlation indicates considerable scatter in the data, which is at least partly due to radio and X-ray variability. This result improves upon previous studies from the unique perspective enabled by VLBA data and shows a positive correlation agreeing with the previous work by \cite{Forbrich:2013} and \cite{Yanza:2022}.

\begin{figure*}
	\includegraphics[width=0.8\paperwidth]{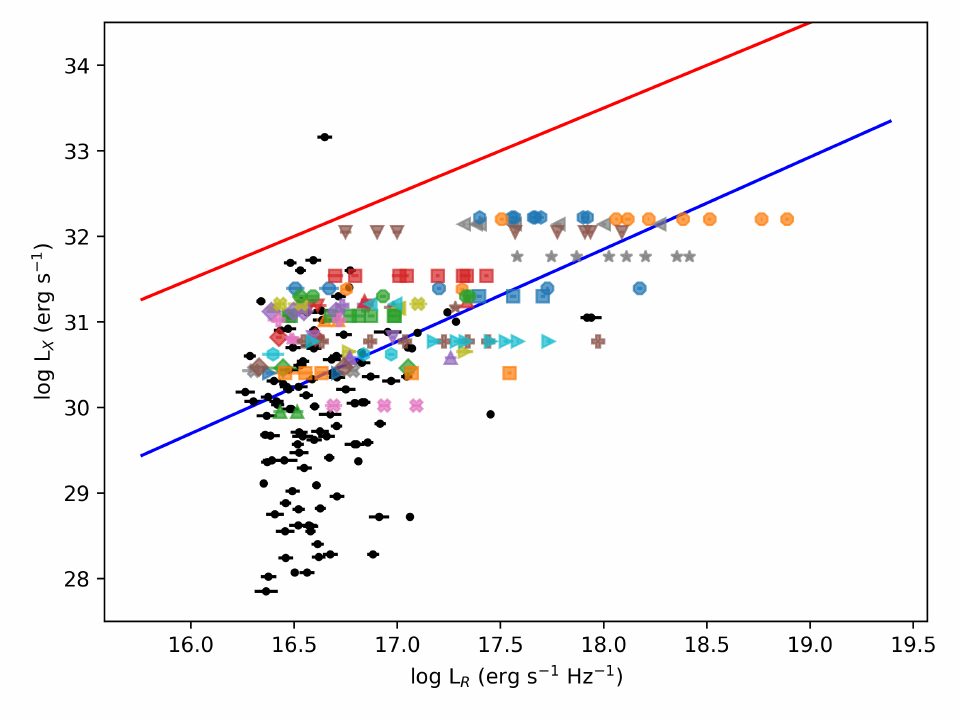}
    \caption{X-ray luminosities as a function of radio luminosity for VLBA detections and associated COUP counterparts. There are 126 VLBA sources that have COUP counterparts, some are detected multiple times totalling 250 detections, the black points are sources detected once and the coloured points with different markers indicate a source detected more than once. This shows the range of radio luminosity measurements for these sources. The blue line is the fit for our data and the red line is the original G$\ddot{\text{u}}$del-Benz relation. The uncertainty in the luminosities are the error propagation of the flux densities and the error in the fit is the error propagation of the luminosities.}
    \label{fig:VLBA_logRXlum}
\end{figure*}

Figure\,\ref{fig:VLBA_logRXlum} shows the X-ray luminosities as a function of the radio luminosities we derived. Sources detected in multiple VLBA epochs are plotted individually using unique markers and colours, and sources detected once are indicated by black markers. This highlights the variability seen in the radio sources. The red line is the original G-B relation and the blue line is our fit. The formal luminosity uncertainties are derived from the propagated flux density errors and the fit uncertainty is the propagated luminosity errors.

Our fit lies roughly two orders of magnitude below the original G-B relation in X-ray luminosities. This has been observed by the other ONC studies previously mentioned, \cite{Forbrich:2013} and \cite{Yanza:2022}, who both found their sources are under-luminous in X-ray (or `radio-bright') compared to the original G-B relation. It should be noted that the work presented by \cite{Yanza:2022} is focused on the M17 nebula but they also investigate this for the ONC. The theoretical YSO modelling work conducted by \cite{Waterfall:2019} also agrees with our results. Their model uses a magnetic flux tube filled with non-thermal electrons and their results show a departure from the G-B relation in a similar manner to our results.

The systematic offset from the original G-B relation indicates that the YSO population occupies a different normalisation. Such offsets are not unexpected for young stars which are highly active and where the radio and X-ray relation is known to differ from magnetically active main-sequence stars \cite[e.g.,][]{Gudel-benz:1993, Forbrich:2013}. In particular, for highly active stars the X-ray emission can reach a saturation level in $L_X / L_{\text{bol}}$ whereas nonthermal radio emission associated with accelerated electrons need not saturate in the same manner \cite[e.g.,][]{Gudel:2004}. This can result in relatively enhanced radio emission and therefore a departure from the canonical G-B relation. The different magnetic and coronal environments of young stars, together with their high levels of magnetic activity, could contribute to the population dependent normalisation. Similarly, \cite{Forbrich:2013} found that nonthermal radio sources and saturated X-ray sources within the ONC are usually closer to the G-B relation but with systematically higher radio luminosities. Additionally, the strong variability of the radio and X-ray emission, combined with the non-simultaneity of the observations, is expected to introduce scatter reducing the statistical significance of the observed offset.

Since the slope in our fit is consistent with unity within the uncertainties, this indicates a nearly linear relation between $L_X$ and $L_R$. In this case the relation can be written approximately in the form of the G-B relation. We can write our fit in the linear form as $L_X/L_R = 10^{12.5 \pm 1.6} L_R^{0.08 \pm 0.10}$. Comparing this expression with the G-B relation (Eq.~\ref{eq:gudel-benz}), we obtain 
\begin{equation}
k = 10^{-3.0} L_R^{0.08} .
\end{equation}
Because the fitted slope is consistent with unity, $k$ depends only weakly on $L_R$. For representative radio luminosities in our sample ($L_R\sim10^{16}$ -- $10^{19}$ erg s$^{-1}$\,Hz$^{-1}$), we obtain $k$ ranging from 0.02 to 0.03, in agreement with the empirical value of $k=0.03$ reported by \citet{Dzib:2015}.


\section{Summary and conclusions}
\label{sect:summary_conclusions}

We present an updated non-thermal census of the Orion Nebula Cluster, building upon previous VLBA studies by \cite{Forbrich2021} and \cite{Dzib2021}. We extend this work with a per-epoch variability analysis. This is based on four more VLBA epochs of follow up observations of the 575 VLA radio sources, totalling eight VLBA epochs each $\sim$7 hours in duration obtained between 2015 and 2021.

The primary goals of this work are to update the non-thermal detection statistics and obtain per-epoch photometry to characterise variability statistics. By measuring epoch averaged peak flux densities we investigate variability on timescales down to 1 day. We find 21 sources that exhibit high variability (VF $\geq$ 5) and that significant variability (where a VF has a S/N $\geq$ 3) is common. The diversity of the observed variability indicates that it is likely occurring on timescales even shorter than our shortest epoch spacing of 1 day. This is consistent with previous studies, for example \cite{Forbrich:2017} found three sources at centimetre wavelengths exhibiting extreme variability on timescales between 0.4 to 0.7 hours and \cite{Vargas-Gonzalez:2024-thesis} reports variability for two sources at high time resolution on the order of hours. \\

\noindent The main results and conclusions can be summarised as:

\begin{itemize}

    \item We present updated non-thermal detection statistics extending the work of \cite{Forbrich2021} by adding four new epochs and analysing detections across all eight epochs. Of the 575 VLA targets we detect 226 unique sources (335 total detections) corresponding to an overall detection fraction of 39 per cent. Although, it is known that some targets are non-stellar in origin and therefore are not expected to be detected by VLBA observations. The majority of the sources appear in only a single epoch (188 sources or $\sim$83 per cent), while 35 sources ($\sim$16 per cent of unique sources) are detected in more than one epoch and of these there are 9 sources ($\sim$4 per cent of unique sources) that are detected in all eight epochs. We report the detection of 93 new sources ($\sim$41 per cent of unique sources) that were not detected in the first four epochs reported by \cite{Forbrich2021}. The predominance of single epoch detections, where we detect a source that is then not detected in subsequent epochs, underlines the highly intermittent nature of the non-thermal source population. Additionally, no significant circularly polarised emission (Stokes $V$) is detected for any source. We estimate upper limits on the circular polarisation fraction that span a wide range of $\sim\!1$ to 100 per cent with a median of $\sim\!13$ per cent.\\

    \item Non-detections in the VLBA data cannot be explained solely by the difference in sensitivity between the VLBA (average rms of 35\,$\mu$Jy beam$^{-1}$) and the VLA (nominal rms of 3\,$\mu$Jy beam$^{-1}$). The absence of VLBA counterparts for a subset of VLA sources could partially be explained by variability but is also consistent with sources lacking a compact non-thermal emission component. \\

    \item Source variability was quantified using the variability factor (VF) derived from \textit{per-epoch} peak flux densities. All derived VFs in this work are considered significant with S/N $\geq$ 3. Of the 226 unique sources 21 ($\sim$9 per cent of unique sources) exhibit high variability (VF $\geq$ 5), including 8 sources ($\sim$4 per cent) showing extreme variability with VF $\geq$ 10. The most variable source in the sample is associated with the class I YSO source ONC130 with a VF of 47.1 $\pm$ 6.4. Significant variability is observed at multiple timescales and in some cases for the same source. The differing VF values at different timescales for individual sources is consistent with variability occurring on timescales shorter than our shortest spacing between epochs of 1 day.\\

    \item Cross-matching our VLBA detections with the spectral classifications from \cite{Hillenbrand:2013} reveals that there are 64 sources that have spectral classifications within our sample. These are mostly M, K and K-M classes, however, it should be highlighted that this only represents $\sim$28 per cent of the VLBA source population. \\

    \item We also cross-matched our VLBA detections with the COUP X-ray survey \citep{Getman:2005} with the main motivation being to identify these sources as YSOs as elevated X-ray emission in this context is associated with pre-main sequence stars. We find 126 of the 226 unique radio sources ($\sim$56 per cent) as having X-ray counterparts. We also cross-match with \textit{Hubble} \citep{Whitmore:2016} and VISION \citep{Meingast:2016} catalogues where we find 138 and 98 associations respectively. There are 36 remaining sources that lack any counterparts. We then estimate the number of background sources within the VLBA primary beam as up to 3 sources and within the imaged area by this work as $\sim$0.2 sources. Therefore, the contribution from background sources is considered negligible and the sources lacking any counterparts are likely ONC members. We also identify 36 newly detected sources that have COUP counterparts confirming these YSOs as having an observable non-thermal radio component and 37 new VLBA detections that do not have COUP counterparts.\\

    \item Investigating the G\"udel-Benz relation, we find that the VLBA detected sources are systematically under-luminous in X-rays relative to the original relation. The offset may be a result of the radio and X-ray observations being non-simultaneous as discussed by \cite{Forbrich:2013}. Despite this offset we find a clear correlation between radio and X-ray luminosities with a positive Pearson correlation coefficient of (R) = 0.58. This is a novel measurement as VLBA observations means the sources are known to be non-thermal and the G-B relation is thought to only hold for non-thermal emission. Whereas, previous work used VLA observations that include both thermal and non-thermal emission \citep[e.g.,][]{Forbrich:2013,Yanza:2022}. \\

    \item We estimate the detectability and variability of YSOs across the eight epochs of this work. Of the 575 targets of this work there are 254 that have COUP counterparts. Of these there are 126 detected in our VLBA epochs totalling 250 detections (as some are detected multiple times). The probability of detecting a YSO in a given epoch is estimated as $\sim$12 per cent, further highlighting that YSOs are highly intermittent. We then considered detections in closely spaced epochs to assess the persistence of radio emission (closely spaced epochs are separated by 1 or 4 days), where we find 19 sources corresponding to a probability of $\sim$9 per cent, showing that emission rarely persists across epochs. We considered the occurrence of higher variability where we find 21 sources ($\sim$8 per cent of detected sources) exhibiting higher variability, of these there are 12 sources ($\sim$5 per cent) that exhibit higher variability between adjacent epochs. Finally, closely spaced epochs (separated by up to a few days) were investigated in the context of high variability where 3 sources exhibit higher variability across closely spaced epochs ($\sim$14 per cent of highly variable sources). These results show that YSO radio emission is highly intermittent and variable, and the low probabilities of detecting a source in a given epoch, and in both adjacent or closely spaced epochs, suggests variability on timescales shorter than the shortest epoch spacing of 1 day.

\end{itemize}

\section*{Acknowledgements}

The National Radio Astronomy Observatory is a facility of the National Science Foundation operated under cooperative agreement by Associated Universities, Inc. S.A.D. acknowledges the M2FINDERS project from the European Research Council (ERC) under the European Union’s Horizon 2020 research and innovation programme (grant No 101018682). E.T.O. acknowledges support from a STFC PhD studentship (grant ST/X508408/1).

\section*{Data Availability}

The VLBA data used in this work (BF117, BF123A, BF123B, BF123C, BF131A, BF131B, BF131C, BF131D) are available from the NRAO data archive (https://data.nrao.edu). The data underlying this article will be shared on reasonable request to the corresponding author.

\begin{landscape}

\begin{table}
	\centering
	\caption{Table of primary beam corrected flux density measurements in mJy beam$^{-1}$ from the resulting \texttt{imfit} \citep{CASA2022} and the inter-epoch variability factors for each VLBA epoch.}
	\label{tab:flux_table}
	\begin{tabular}{lcccccccccc} 
		\hline
        \hline
		[FRM2016] & COUP & BF117           & BF123A           & BF123B          & BF123C          & BF131A       & BF131B       & BF131C       & BF131D & Variability Factor\\
                  &      & [2015-10-26]    & [2017-10-26]     & [2017-10-27]    & [2018-10-26]    & [2020-10-25] & [2020-10-29] & [2021-10-30] & [2021-10-31] \\
		\hline
        002.1*     & 107  & -               &  -               & 0.29 $\pm$ 0.02 & -               & -            & -            & -            & - & >1.2\\
        002.2*     & 107  & -               &  -               & -               & 2.80 $\pm$ 0.07 & 2.01 $\pm$ 0.07 & 1.38 $\pm$ 0.04 & -      & - & 2.0 $\pm$ 0.1\\
        003.1*     & -    & - & - & - & - & - & - & 0.75 $\pm$ 0.08 & - & >6.1\\
        003.2*     & -    & - & - & - & - & - & - & - & 0.45 $\pm$ 0.06 & >3.7\\
        004       & 141  & - & - & - & - & - & - & - & 0.37 $\pm$ 0.03 & >2.1\\
        005       & 172  & - & - & - & - & - & - & 0.41 $\pm$ 0.04 & - & >2.2\\
        010.1     & 262  & -               &  -               & -               & 0.29 $\pm$ 0.01 & - & 0.25 $\pm$ 0.02 & -      & - & 1.2 $\pm$ 0.1\\
        010.2     & 262  & -               &  -               & -               & - & 0.24 $\pm$ 0.02 & - & -      & - & >2.0\\
        011       & -    & -               &  -               & 0.26 $\pm$ 0.01 & -               & -            & -            & -            & - & >1.5\\
        014.1     & 283  & - & 0.17 $\pm$ 0.02 & - & - & - & - & - & - & >1.4\\
        014.2     & 283  & - & - & - & - & - & - & - & 0.17 $\pm$ 0.01 & >1.4\\
        016       & 322  & -               & -                & -               & -             & 1.56 $\pm$ 0.03 & -           & -            & - & >12.5\\
        018       & 338  & 0.62 $\pm$ 0.03 &  0.15 $\pm$ 0.03 & -               & - & - & - & - & - & 3.9 $\pm$ 0.6            \\
        022       & 342  & 0.23 $\pm$ 0.02               &  -               & 0.40 $\pm$ 0.03 & -             & 0.21 $\pm$ 0.01 & - & - & - & 1.9 $\pm$ 0.2\\
        024       & 350  & 0.19 $\pm$ 0.01 &  -               & -               & - & - & - & - & - & >1.4            \\
        030       & 363  & -               &  0.16 $\pm$ 0.02 & -               & - & - & - & - & - & >1.3            \\
        035       & 390  & -               &  -               & -               & -             & 0.28 $\pm$ 0.02 & - & - & - & >2.3\\
        037       & 394  & -               &  -               & 0.19 $\pm$ 0.01 & -             & - & - & 0.38 $\pm$ 0.02 & 0.30 $\pm$ 0.01 & 1.9 $\pm$ 0.2\\
        042       & -    & -               &  -               & -               & 0.17 $\pm$ 0.01 & - & - & - & - & >1.2\\
        051*       & 430  & -               &  -               & -               & -             & 1.06 $\pm$ 0.05 & 0.53 $\pm$ 0.05 & - & - & 1.9 $\pm$ 0.2\\
        053       & 427  & -               &  -               & -               & 0.37 $\pm$ 0.02 & - & - & - & - & >1.8\\
        055       & -    & -               &  -               & 0.24 $\pm$ 0.02 & - & - & - & - & - & >1.4\\
        059       & 434  & - & - & - & - & - & - & - & 0.16 $\pm$ 0.01 & >1.1\\
        064       & 444  & 0.48 $\pm$ 0.03 &  - & 0.27 $\pm$ 0.03 & - & - & 0.68 $\pm$ 0.04 & - & - & 2.5 $\pm$ 0.2\\
	    066       & 450  & 1.33 $\pm$ 0.05 &  1.99 $\pm$ 0.03 & 1.16 $\pm$ 0.02 & 1.39 $\pm$ 0.02 & 5.55 $\pm$ 0.11 & 2.05 $\pm$ 0.06 & 10.29 $\pm$ 0.18 & 3.34 $\pm$ 0.12 & 8.8 $\pm$ 0.2\\
        068       & -    & - & - & - & - & - & - & 0.18 $\pm$ 0.02 & - & >1.5\\
        070       & 454  & -               &  -               & 0.25 $\pm$ 0.02 & - & - & - & - & - & >1.6\\
        072       & -    & -               &  -               & 0.15 $\pm$ 0.01 & - & - & - & - & - & >1.3\\
        075.1     & 465  & -               &  -               & -               & 0.18 $\pm$ 0.01 & - & - & - & - & >1.5\\
        075.2     & 465  & -               &  -               & -               & - & - & - & - & 0.20 $\pm$ 0.02 & >1.7\\
        079       & 476  & - & - & - & - & - & - & - & 0.13 $\pm$ 0.01 & >1.1\\
        091       & -    & - & - & - & - & - & - & - & 0.16 $\pm$ 0.02 & >1.3\\
        093       & 504  & 1.17 $\pm$ 0.11 &  -               & -               & - & - & - & 0.32 $\pm$ 0.02 & - & 3.6 $\pm$ 0.4\\
        098       & 510  & 0.24 $\pm$ 0.02 &  -               & -               & - & - & - & - & - & >1.7\\
        102       & -    & - & - & - & - & - & - & 0.29 $\pm$ 0.03 & - & >1.9\\
        105       & -    & - & - & - & - & - & 0.22 $\pm$ 0.02 & - & - & >1.5\\
        110       & 530  & -               &  -               & -               & -               & 0.52 $\pm$ 0.02 & - & 0.38 $\pm$ 0.02 & 0.14 $\pm$ 0.02 & 3.6 $\pm$ 0.5\\
        122       & -    & -               &  -               & 0.13 $\pm$ 0.02 & - & - & - & - & - & >1.1\\
        127.1     & 551  & -               &  -               & 0.14 $\pm$ 0.01 & - & - & - & - & - & >1.2\\
        127.2     & 551  & -               &  -               & - & - & - & - & 0.16 $\pm$ 0.02 & - & >1.4\\
        129       & -    & -               &  -               & 0.18 $\pm$ 0.01 & - & - & - & - & - & >1.3\\

		\hline
	\end{tabular}
\end{table}

\end{landscape}


\begin{landscape}
\newpage
\begin{table}
	\centering
        \contcaption{}
	
	\begin{tabular}{lcccccccccc} 
		\hline
        \hline
		[FRM2016] & COUP & BF117           & BF123A           & BF123B          & BF123C          & BF131A       & BF131B       & BF131C       & BF131D & Variability Factor\\
                  &      & [2015-10-26]    & [2017-10-26]     & [2017-10-27]    & [2018-10-26]    & [2020-10-25] & [2020-10-29] & [2021-10-30] & [2021-10-31] \\
		\hline
        130       & 554  & 0.26 $\pm$ 0.03 &  0.88 $\pm$ 0.02 & 2.95 $\pm$ 0.05 & 8.27 $\pm$ 0.22 & 0.18 $\pm$ 0.02 & - & - & - & 47.1 $\pm$ 6.4\\
        133.1     & -    & -               &  -               & 0.17 $\pm$ 0.02 & - & - & - & - & - & >1.5\\
        133.2     & -    & -               &  -               & - & - & - & - & 0.31 $\pm$ 0.02 & - & >2.7\\
        135       & 571  & -               &  0.16 $\pm$ 0.01 & -               & - & - & - & - & - & >1.3\\
        137       & -    & -               &  -               & 0.23 $\pm$ 0.02 & 0.19 $\pm$ 0.02 & - & - & - & - & 1.2 $\pm$ 0.1\\
        148       & 593  & -               &  0.17 $\pm$ 0.01 & -               & - & - & - & - & - & >1.4\\
        154       & 594  & 0.65 $\pm$ 0.02 &  -               & 0.24 $\pm$ 0.01 & 1.94 $\pm$ 0.04 & 0.20 $\pm$ 0.02 & - & - & 0.16 $\pm$ 0.02 & 12.0 $\pm$ 1.9\\
        158       & 602  & -               &  0.18 $\pm$ 0.01 & -               & 0.15 $\pm$ 0.01 & - & - & - & - & 1.1 $\pm$ 0.1\\
        161       & 598  & 0.26 $\pm$ 0.03 &  -               & -               & - & - & - & - & - & >1.5\\
        167*       & 608  & -               &  -               & -               & 0.46 $\pm$ 0.03 & - & - & - & - & >1.0\\
        170       & 622  & -               &  -               & -               & 0.13 $\pm$ 0.01 & - & - & - & - & >1.0\\
        176       & -    & -               &  -               & -               & 0.14 $\pm$ 0.01 & - & - & - & - & >1.0\\
        177.1     & 625  & 0.22 $\pm$ 0.02 &  -               & -               & 0.15 $\pm$ 0.01 & - & - & - & - & 1.5 $\pm$ 0.2\\
        177.2     & 625  & -               &  -               & -               & 0.53 $\pm$ 0.01 & - & - & - & 0.22 $\pm$ 0.01 & 2.4 $\pm$ 0.2\\
        182       & -    & -               &  -               & 0.19 $\pm$ 0.01 & - & - & - & - & - & >1.5\\
        184       & 639  & 0.19 $\pm$ 0.02 &  0.41 $\pm$ 0.02 & 0.24 $\pm$ 0.01 & 0.93 $\pm$ 0.02 & 5.19 $\pm$ 0.08 & 1.53 $\pm$ 0.04 & 1.21 $\pm$ 0.02 & 0.60 $\pm$ 0.03 & 27.3 $\pm$ 2.9\\
        188       & -    & -               &  0.17 $\pm$ 0.01 & -               & - & - & - & - & - & >1.4\\
        189       & 640  & 0.20 $\pm$ 0.02 &  -               & -               & - & - & - & - & - & >1.7\\
        193       & -    & - & - & - & - & - & - & 0.18 $\pm$ 0.02 & - & >1.6\\
        194       & -    & - & - & - & - & - & - & - & 0.14 $\pm$ 0.01 & >1.2\\
        196       & 648  & 0.69 $\pm$ 0.03 &  -               & -               & - & - & - & - & - & >6.0\\
        197       & 645  & -               &  0.19 $\pm$ 0.01 & -               & - & - & - & - & - & >1.3\\
        198       & 647  & -               &  -               & 0.17 $\pm$ 0.02 & 0.17 $\pm$ 0.01 & - & - & - & - & 1.0 $\pm$ 0.1\\
        199       & 649  & - & - & - & - & 0.34 $\pm$ 0.02 & - & - & 0.11 $\pm$ 0.01 & 3.0 $\pm$ 0.3 \\
        203       & 655  & - & - & - & - & - & 0.19 $\pm$ 0.01 & - & - & >1.6\\
        205       & -    & 0.17 $\pm$ 0.02 &  -               & -               & - & - & - & - & - & >1.5\\
        211       & 662  & 0.19 $\pm$ 0.02 &  0.56 $\pm$ 0.01 & 0.56 $\pm$ 0.02 & 0.14 $\pm$ 0.01 & - & - & - & - & 4.0 $\pm$ 0.3\\
        212       & 670  & 1.06 $\pm$ 0.02 &  -               & -               & - & - & - & - & - & >9.2\\
        219       & 680  & - & - & - & - & - & - & - & 0.62 $\pm$ 0.03 & >5.3\\
        222       & 689  & -               &  -               & 0.17 $\pm$ 0.01 & - & - & - & - & - & >1.4\\
        227       & -    & -               &  -               & 0.13 $\pm$ 0.01 & - & - & - & - & - & >1.0\\
        229       & -    & 0.21 $\pm$ 0.02 & - & - & - & - & - & - & - & >1.7\\
        230       & -    & -               &  -               & -               & 0.12 $\pm$ 0.01 & - & - & - & - & >1.0\\
        232       & -    & -               &  -               & -               & 0.13 $\pm$ 0.01 & - & - & - & - & >1.0\\
        240       & 717  & 0.23 $\pm$ 0.01 &  -               & -               & - & - & - & - & - & >1.9\\
        241       & 718  & 1.57 $\pm$ 0.07 &  1.18 $\pm$ 0.02 & 0.22 $\pm$ 0.02 & 2.12 $\pm$ 0.03 & 2.99 $\pm$ 0.07 & 0.83 $\pm$ 0.05 & 1.86 $\pm$ 0.05 & 1.08 $\pm$ 0.04 & 13.5 $\pm$ 1.2\\
        243       & -    & - & - & - & - & - & - & - & 0.12 $\pm$ 0.01 & >1.0\\
        244.1     & 724  & - & - & - & - & - & 0.33 $\pm$ 0.02 & - & - & >2.8\\
        244.2     & 724  & - & - & - & - & - & - & 0.19 $\pm$ 0.01 & - & >1.6\\
        \hline
	\end{tabular}
\end{table}

\end{landscape}

\begin{landscape}
\newpage
\begin{table}
	\centering
        \contcaption{}
	
	\begin{tabular}{lcccccccccc} 
		\hline
        \hline
		[FRM2016] & COUP & BF117           & BF123A           & BF123B          & BF123C          & BF131A       & BF131B       & BF131C       & BF131D & Variability Factor\\
                  &      & [2015-10-26]    & [2017-10-26]     & [2017-10-27]    & [2018-10-26]    & [2020-10-25] & [2020-10-29] & [2021-10-30] & [2021-10-31] \\
		\hline
        249       & 734  & -               &  -               & -               & 0.13 $\pm$ 0.01 & - & - & - & - & >1.0\\
        250       & 732  & 1.99 $\pm$ 0.02 &  1.39 $\pm$ 0.02 & 2.58 $\pm$ 0.04 & 2.52 $\pm$ 0.05 & 4.40 $\pm$ 0.10 & 2.05 $\pm$ 0.07 & 4.62 $\pm$ 0.07 & 2.74 $\pm$ 0.09 & 3.3 $\pm$ 0.1\\
        254       & 745  & 42.72 $\pm$ 0.87 &  9.14 $\pm$ 0.08 & 1.77 $\pm$ 0.04 & 32.23 $\pm$ 0.32 & 13.42 $\pm$ 0.26 & 18.03 $\pm$ 0.32 & 6.37 $\pm$ 0.09 & 7.23 $\pm$ 0.25 & 24.0 $\pm$ 0.6\\
        260       & -    & - & - & - & - & - & 0.17 $\pm$ 0.02 & - & - & >1.1\\
        267       & 763  & 0.20 $\pm$ 0.02 & - & - & - & - & - & - & - & >1.6\\
        268       & 762  & -               &  -                & -              & - & - & - & - & 0.16 $\pm$ 0.02 & >1.2\\
        270       & 760  & -               & -                 &                & - & 0.28 $\pm$ 0.02 & - & - & - & >1.3\\
        273       & -    & -               &  -                & -              & 0.15 $\pm$ 0.01 & - & - & - & - & >1.0\\
        274       & 765  & - & - & - & - & - & - & - & 0.14 $\pm$ 0.01 & >1.2\\
        300.1     & 801  & 0.20 $\pm$ 0.02 &  -                & -              & - & - & - & - & - & >1.7\\
        300.2     & 801  & - &  -                & -              & - & - & 0.97 $\pm$ 0.04 & - & - & >8.2\\
        303       & 806  & -               &  -                & 0.15 $\pm$ 0.01 & - & - & - & - & - & >1.2\\
        305       & 809  & - & - & - & - & - & - & - & 0.24 $\pm$ 0.02 & >2.0\\ 
        314       & -    & 0.30 $\pm$ 0.03 &  -                & -               & - & - & - & - & - & >1.9\\
        319       & 828  & 0.54 $\pm$ 0.04 &  0.33 $\pm$ 0.02  & 0.17 $\pm$ 0.01 & 0.16 $\pm$ 0.01 & 0.41 $\pm$ 0.01 & 0.26 $\pm$ 0.02 & 0.53 $\pm$ 0.02 & 0.36 $\pm$ 0.02 & 3.4 $\pm$ 0.3\\
        321       & 827  & -               &  0.21 $\pm$ 0.01  & -               & - & - & - & - & - & >1.7\\
        325       & 847 & - & - & - & - & - & - & 0.22 $\pm$ 0.01 & - & >1.7\\
        326       & 841  & -               &  -                & 0.21 $\pm$ 0.01 & - & - & - & - & - & >1.8\\
        327       & 855    & 0.21 $\pm$ 0.02 &  -                & -               & - & - & - & - & - & >1.6\\
        334       & 854  & -               &  -                & -               & - & 1.20 $\pm$ 0.02 & 0.38 $\pm$ 0.02 & - & - & 3.1 $\pm$ 0.2\\
        335       & 856  & -               &  0.16 $\pm$ 0.01  & -               & - & - & - & - & - & >1.3\\
        339.1     & -    & -               &  -                & -               & 0.17 $\pm$ 0.01 & - & - & - & - & >1.4\\
        339.2     & -    & -               &  -                & -               & - & - & - & 0.31 $\pm$ 0.03 & - & >2.7\\
        340       & -    & -               &  -                & -               & - & - & 0.16 $\pm$ 0.02 & - & - & >1.3\\
        343       & 867  & 0.22 $\pm$ 0.02 &  -                & -               & - & - & - & - & - & >1.7\\
        344       & -    & -               & -                 & -               & - & - & 0.19 $\pm$ 0.02 & - & - & >1.5\\
        347       & 876  & -               &  -                & -               & 0.13 $\pm$ 0.01 & - & - & - & - & >1.0\\
        350       & 874  & 0.20 $\pm$ 0.01 &  0.14 $\pm$ 0.01  & -               & 0.28 $\pm$ 0.01 & - & 0.17 $\pm$ 0.02 & - & - & 1.6 $\pm$ 0.2\\
        360       & 897  & -               &  -                & 0.14 $\pm$ 0.01 & - & - & - & - & - & >1.2\\
        364       & 912    & -               &  -                & 0.20 $\pm$ 0.01 & - & - & - & - & - & >1.5\\
        367       & -    & -               &  0.23 $\pm$ 0.01  & -               & - & - & - & - & - & >1.3\\
        377*       & -    & - & - & - & - & - & - & - & 0.28 $\pm$ 0.03 & >1.3\\
        378       & 932  & 6.77 $\pm$ 0.15 &  3.30 $\pm$ 0.06  & 2.06 $\pm$ 0.03 & 4.48 $\pm$ 0.09 & 0.31 $\pm$ 0.02 & 4.76 $\pm$ 0.15 & 0.55 $\pm$ 0.03 & 0.44 $\pm$ 0.02 & 22.0 $\pm$ 1.5\\
        382       & 942  & 0.15 $\pm$ 0.02 &  -                & -               & 0.29 $\pm$ 0.02 & - & - & - & - & 2.0 $\pm$ 0.3\\
        389       & 956  & 0.18 $\pm$ 0.02 &  -                & -               & - & - & - & - & - & >1.4\\
        398       & -    & -               &  -                & -               & 0.23 $\pm$ 0.02 & - & - & - & - & >1.0\\
        400       & 965  & 5.86 $\pm$ 0.14 &  2.11 $\pm$ 0.06  & 3.08 $\pm$ 0.06 & 14.40 $\pm$ 0.32 & 4.09 $\pm$ 0.08 & 12.53 $\pm$ 0.37 & 7.13 $\pm$ 0.12 & 8.83 $\pm$ 0.30 & 6.8 $\pm$ 0.3\\
        402       & 963  & -               &  -                & -               & 0.18 $\pm$ 0.02 & - & - & - & - & >1.0\\
        408       & -    & -               &  -                & 0.16 $\pm$ 0.02 & - & - & - & - & - & >1.2\\
        414.1     & 985  & 0.70 $\pm$ 0.07 &  0.19 $\pm$ 0.03  & -               & 0.15 $\pm$ 0.01 & - & - & - & - & 4.5 $\pm$ 0.6\\
        414.2     & 985  & -               &  -                & 0.41 $\pm$ 0.02 & 0.54 $\pm$ 0.02 & - & - & - & - & 1.3 $\pm$ 0.1\\
        \hline
	\end{tabular}
\end{table}

\end{landscape}

\begin{landscape}
\newpage
\begin{table}
	\centering
        \contcaption{}
	
	\begin{tabular}{lcccccccccc} 
		\hline
        \hline
		[FRM2016] & COUP & BF117           & BF123A           & BF123B          & BF123C          & BF131A       & BF131B       & BF131C       & BF131D & Variability Factor\\
                  &      & [2015-10-26]    & [2017-10-26]     & [2017-10-27]    & [2018-10-26]    & [2020-10-25] & [2020-10-29] & [2021-10-30] & [2021-10-31] \\
		\hline
        418       & -    & - & - & - & - & - & - & - & 0.16 $\pm$ 0.01 & >1.2\\
        419       & -    & - & - & - & - & - & - & 0.28 $\pm$ 0.02 & - & >2.1\\
        426*       & -    & -               &  -                & 0.25 $\pm$ 0.03 & - & - & - & - & - & >1.2\\
        427.1     & 1008 & - & - & - & - & - & - & 0.18 $\pm$ 0.02 & - & >1.5\\
        427.2     & 1008 & - & - & - & - & - & - & 0.16 $\pm$ 0.01 & - & >1.3\\
        430       & -    & - & - & - & - & - & - & 0.26 $\pm$ 0.03 & - & >1.9\\
        434       & 1016 & - & - & - & - & - & - & 0.28 $\pm$ 0.02 & - & >2.1\\
        435       & 1023  & -               &  -                & 0.29 $\pm$ 0.02 & - & 0.13 $\pm$ 0.01 & - & - & - & 2.1 $\pm$ 0.2\\
        440       & -    & -               &  -                & -               & 0.14 $\pm$ 0.01 & - & - & - & - & >1.0\\
        441       & -    & - & - & - & - & - & - & 0.29 $\pm$ 0.02 & - & >2.2\\
        444.1*     & 1035 & -               & -                 & -               & - & - & 4.82 $\pm$ 0.55 & - & - & >32.4\\
        444.2*     & 1035 & -               & -                 & -               & - & - & - & - & 4.58 $\pm$ 0.32 & >30.8\\
        452       & -    & - & - & - & - & - & - & 0.26 $\pm$ 0.02 & - & >2.1\\
        453       & 1056 & - & - & - & - & - & - & 0.22 $\pm$ 0.02 & - & >1.8\\
        456       & -    & -               &  -                & 0.18 $\pm$ 0.01 & - & - & - & - & - & >1.3\\
        457       & 1071  & -               &  0.17 $\pm$ 0.01  & -               & - & - & - & - & - & >1.2\\
        459       & 1083  & 0.65 $\pm$ 0.02 &  -                & -               & - & - & - & - & - & >4.9\\
        462*       & 1087  & -               &  -                & -               & 0.32 $\pm$ 0.01 & - & - & - & - & >2.5\\
        466       & 1090  & -               &  0.31 $\pm$ 0.01  & 1.14 $\pm$ 0.03 & - & - & - & - & - & 3.7 $\pm$ 0.2\\
        467*       & 1091  & -               &  -                & 0.26 $\pm$ 0.01 & - & - & - & - & - & >1.2\\
        468       & 1100  & -               &  -                & -               & 0.22 $\pm$ 0.01 & - & - & - & - & >1.6\\
        469       & 1101  & - & - & - & - & - & - & - & 0.19 $\pm$ 0.01 & >1.4\\
        470       & -     & -               &  -                & -               & 0.17 $\pm$ 0.01 & - & - & - & - & >1.0\\
        476       & -     & - & - & - & - & 0.21 $\pm$ 0.02 & - & - & - & >1.6\\
        477       & 1110  & -               &  -                & -               & 0.13 $\pm$ 0.01 & - & - & - & - & >1.0\\
        478*       & -     & - & - & - & - & - & - & 0.34 $\pm$ 0.03 & - & >1.7\\
        480       & 1116  & 0.22 $\pm$ 0.01 &  -                & -               & 0.47 $\pm$ 0.02 & 0.19 $\pm$ 0.02 & - & 1.24 $\pm$ 0.04 & 1.19 $\pm$ 0.04 & 6.5 $\pm$ 0.4\\
        485       & 1130  & 1.15 $\pm$ 0.05 &  0.35 $\pm$ 0.01  & 0.87 $\pm$ 0.02 & 0.61 $\pm$ 0.01 & 1.50 $\pm$ 0.03 & 1.20 $\pm$ 0.05 & 0.57 $\pm$ 0.05 & 0.28 $\pm$ 0.02 & 5.5 $\pm$ 0.5\\
        486*       & 1139  & - & - & - & - & - & - & - & 0.52 $\pm$ 0.05 & >1.5\\
        493*       & 1151  & - & - & - & - & - & - & 0.21 $\pm$ 0.02 & - & >1.8\\
        501       & 1184  & -               &  0.23 $\pm$ 0.01  & -               & - & - & - & - & - & >1.3\\
        507       & -     & - & - & - & - & - & - & 0.25 $\pm$ 0.01 & - & >1.7\\
        508       & 1205  & -               &  -                & -               & 0.18 $\pm$ 0.01 & - & - & - & - & >1.0\\
        509*       & -    & -               &  -                & -               & 0.25 $\pm$ 0.01 & - & - & - & - & >1.0\\
        512       & -    & -               &  -                & 0.17 $\pm$ 0.01 & - & - & - & - & - & >1.3\\
        513*       & 1224 & -               & -                 & -               & - & 0.61 $\pm$ 0.03 & - & - & - & >2.6\\
        514*       & -    & -               &  -                & 0.43 $\pm$ 0.02 & - & - & - & - & - & >1.3\\
        515*       & 1232  & -               &  -                & -               & 0.22 $\pm$ 0.02 & - & - & - & - & >1.0\\
        519*       & -    & - & - & - & - & - & - & 0.41 $\pm$ 0.03 & - & >1.9\\
        520*       & 1249  & -               &  -                & -               & 0.32 $\pm$ 0.02 & - & - & 1.00 $\pm$ 0.06 & 0.33 $\pm$ 0.02 & 3.1 $\pm$ 0.3\\
        521.1*     & 1262  & 0.45 $\pm$ 0.05 &  -                & -               & - & - & - & - & - & >3.7\\
        521.2*     & 1262  & - &  -                & -               & - & - & 0.64 $\pm$ 0.02 & - & - & >5.2\\
        \hline
	\end{tabular}
\end{table}

\end{landscape}

\begin{landscape}
\newpage
\begin{table}
	\centering
        \contcaption{}
	
	\begin{tabular}{lcccccccccc} 
		\hline
        \hline
		[FRM2016] & COUP & BF117           & BF123A           & BF123B          & BF123C          & BF131A       & BF131B       & BF131C       & BF131D & Variability Factor\\
                  &      & [2015-10-26]    & [2017-10-26]     & [2017-10-27]    & [2018-10-26]    & [2020-10-25] & [2020-10-29] & [2021-10-30] & [2021-10-31] \\
		\hline
        522       & 1260  & -               &  -                & 0.28 $\pm$ 0.02 & - & - & - & - & - & >1.8\\
        525.1*     & 1275  & -               &  0.34 $\pm$ 0.04  & -               & - & - & - & - & - & >2.8\\
        525.2*     & 1275  & -               &  -  & -               & - & - & - & - & 0.35 $\pm$ 0.02 & >2.9\\
        526       & 1276  & 0.23 $\pm$ 0.02 &  -                & -               & - & - & - & - & - & >1.6\\
        527*       & 1281  & -               &  -                & 0.40 $\pm$ 0.02 & - & - & - & - & - & >1.4\\
        529*       & 1290 & -                & -                 & -               & -             & 0.27 $\pm$ 0.03 & - & - & - & >1.3\\
        530*       & 1289  & -               &  -                & -               & 0.18 $\pm$ 0.02 & - & - & - & - & >1.0\\
        534*       & -    & -               &  -                & 0.29 $\pm$ 0.02 & - & - & - & - & - & >1.7\\
        535*       & 1313  & -               &  -                & -               & 0.23 $\pm$ 0.02 & - & - & - & - & >1.0\\
        536       & 1309  & - & - & - & - & - & - & - & 0.17 $\pm$ 0.01 & >1.2\\
        537.1*     & 1311  & -               &  0.38 $\pm$ 0.02  & -               & - & - & - & - & - & >3.2\\
        537.2*     & 1311  & -               &  -  & -               & - & - & - & 0.37 $\pm$ 0.02 & - & >3.2\\
        538.1     & 1314  & - & 0.25 $\pm$ 0.02 & - & - & - & - & - & - & >2.1\\
        538.2     & 1314  & - & - & - & - & 0.19 $\pm$ 0.02 & - & - & - & >1.6\\
        539       & -     & - & - & - & - & - & - & 0.33 $\pm$ 0.03 & - & >2.3\\
        545*       & 1341  & - & - & - & - & - & - & 0.30 $\pm$ 0.03 & - & >1.9\\
        546*       & 1350  & - & - & - & - & - & - & - & 0.22 $\pm$ 0.01 & >1.9\\
        547*       & 1360  & -               &  -                & 0.36 $\pm$ 0.02 & - & - & - & - & - & >3.1\\
        548.1*     & 1359  & 0.34 $\pm$ 0.02  & - & - & - & - & - & - & - & >2.9\\
        548.2*     & 1359  & -  & - & - & - & 0.34 $\pm$ 0.03 & - & - & - & >2.9\\
        551*       & - & - & - & - & - & - & - & - & 0.39 $\pm$ 0.03 & >2.0\\
        552.1*     & 1428  & -               &  -                & 0.26 $\pm$ 0.02 & - & - & - & - & - & >2.2\\
        552.2*     & 1428  & -               &  -                & - & - & - & - & 0.42 $\pm$ 0.03 & - & >3.5\\
        555*       & 1473  & -               &  0.50 $\pm$ 0.08  & -               & - & - & - & - & - & >4.2\\
        556*       & 1550  & - & - & - & - & - & - & 0.31 $\pm$ 0.03 & - & >2.7\\
        557       & 672  & -               &  -                & 0.30 $\pm$ 0.02 & 0.12 $\pm$ 0.01 & - & - & - & - & 2.5 $\pm$ 0.3\\
        574       & -    & - & - & - & - & - & - & 0.21 $\pm$ 0.02 & - & >1.6\\
		
        \hline
	\end{tabular}
\raggedright \\
{\small NOTE - columns are left to right: FRM2016 source number, COUP number, epoch names and dates and the variability factor (where they are quoted as greater than represents a lower limit).} \\
{\small * Source is outside the $\sim$80 per cent power point of the FWHM and the uncertainties might be underestimated.}
\end{table}

\end{landscape}

\newpage

\begin{landscape}

\begin{table}
	\centering
	\caption{Table of source detections for all eight VLBA epochs.}
	\label{tab:source_detections}
	\begin{tabular}{lccccccccccccc} 
		\hline
        \hline
		[FRM2016] & Epoch & COUP & Flux Density & S/N & $\alpha_{\text{J}2000}$ & $\sigma_{\alpha}$ & $\delta_{\text{J}2000}$ & $\sigma_{\delta}$ & $\Delta\theta$ & YSO Class & SpType\\
         & & & (mJy beam$^{-1}$) & & & (mas) & & (mas) & (mas) \\
		\hline
ONC002.1* &  3 & 107  & 0.29 $\pm$ 0.02 & 12.3 & 05:34:55.975453 & 0.144 & -05:23:13.02408 & 0.237 & 156.8 $\pm$ 0.4 & III & K1-K4 \\
ONC002.2* &  4 & 107  & 2.80 $\pm$ 0.07 & 41.1 & 05:34:55.974463 & 0.110 & -05:23:13.03819 & 0.120 & 173.0 $\pm$ 0.1 & III & K1-K4 \\
ONC002.2* &  5 & 107  & 2.02 $\pm$ 0.07 & 27.9 & 05:34:55.974645 & 0.120 & -05:23:13.03824 & 0.125 & 172.6 $\pm$ 0.1 & III & K1-K4 \\
ONC002.2* &  6 & 107  & 1.38 $\pm$ 0.04 & 34.5 & 05:34:55.974669 & 0.117 & -05:23:13.03836 & 0.125 & 172.6 $\pm$ 0.1 & III & K1-K4 \\
ONC003.1* &  7 & ---  & 0.75 $\pm$ 0.08 & 8.9 & 05:34:56.702516 & 0.163 & -05:20:33.40197 & 0.194 & 159.6 $\pm$ 0.6 &   &  \\
ONC003.2* &  8 & ---  & 0.45 $\pm$ 0.06 & 7.4 & 05:34:56.687174 & 0.280 & -05:20:32.92400 & 0.423 & 441.1 $\pm$ 6.8 &   &  \\
ONC004*   &  8 & 141  & 0.37 $\pm$ 0.04 & 10.5 & 05:35:00.087741 & 0.129 & -05:23:02.12618 & 0.233 & 274.5 $\pm$ 0.4 & II  &  \\
ONC005*   &  7 & 172  & 0.41 $\pm$ 0.04 & 10.2 & 05:35:02.033498 & 0.124 & -05:20:55.59794 & 0.221 & 567.8 $\pm$ 0.5 &   &  \\
ONC010.1  &  4 & 262  & 0.29 $\pm$ 0.01 & 27.9 & 05:35:06.283553 & 0.111 & -05:22:02.66559 & 0.129 & 23.3 $\pm$ 0.1 & III & K5 \\
ONC010.2  &  5 & 262  & 0.24 $\pm$ 0.02 & 9.6 & 05:35:06.280761 & 0.125 & -05:22:02.73416 & 0.224 & 96.5 $\pm$ 0.1 & III & K5 \\
ONC010.1  &  6 & 262  & 0.25 $\pm$ 0.02 & 10.8 & 05:35:06.283604 & 0.116 & -05:22:02.66651 & 0.187 & 24.1 $\pm$ 0.1 & III & K5 \\
ONC011    &  3 & ---  & 0.26 $\pm$ 0.01 & 22.7 & 05:35:06.416899 & 0.123 & -05:24:21.34737 & 0.182 & 173.9 $\pm$ 0.2 &   &  \\
ONC014.1  &  2 & 283  & 0.17 $\pm$ 0.02 & 10.3 & 05:35:07.243905 & 0.154 & -05:22:26.30108 & 0.358 & 378.2 $\pm$ 1.1 &     & M4.5 \\
ONC014.2  &  8 & 283  & 0.17 $\pm$ 0.01 & 15.1 & 05:35:07.274686 & 0.111 & -05:22:26.67022 & 0.149 & 117.8 $\pm$ 0.1 &     & M4.5 \\
ONC016    &  5 & 322  & 1.56 $\pm$ 0.03 & 52.2 & 05:35:08.736869 & 0.110 & -05:22:56.69107 & 0.114 & 19.8 $\pm$ 0.1 &     & M1.5 \\
ONC018    &  1 & 338  & 0.62 $\pm$ 0.03 & 20.7 & 05:35:09.675505 & 0.123 & -05:23:55.91207 & 0.176 & 41.8 $\pm$ 0.1 &   &  \\
ONC018    &  2 & 338  & 0.15 $\pm$ 0.03 & 5.4 & 05:35:09.676613 & 0.203 & -05:23:55.93909 & 0.448 & 67.7 $\pm$ 0.5 &   &  \\
ONC021    &  A & 343  & 0.15 $\pm$ 0.01 & 13.2 & 05:35:09.769718 & 0.127 & -05:23:26.88885 & 0.276 & 7.5 $\pm$ 0.1 & III & K4-M0 \\
ONC022    &  1 & 342  & 0.23 $\pm$ 0.02 & 10.4 & 05:35:09.770302 & 0.149 & -05:21:28.34616 & 0.269 & 9.5 $\pm$ 0.1 &     & M0, K5-K7 \\
ONC022    &  3 & 342  & 0.40 $\pm$ 0.03 & 15.5 & 05:35:09.769963 & 0.121 & -05:21:28.34776 & 0.170 & 11.5 $\pm$ 0.1 &     & M0, K5-K7 \\
ONC022    &  5 & 342  & 0.21 $\pm$ 0.01 & 14.1 & 05:35:09.769547 & 0.113 & -05:21:28.35828 & 0.163 & 22.7 $\pm$ 0.1 &     & M0, K5-K7 \\
ONC024    &  1 & 350  & 0.18 $\pm$ 0.01 & 14.8 & 05:35:09.882650 & 0.128 & -05:23:38.33139 & 0.225 & 215.0 $\pm$ 0.3 &   &  \\
ONC025    &  A & ---  & 0.11 $\pm$ 0.01 & 11.7 & 05:35:10.043950 & 0.126 & -05:21:21.93646 & 0.256 & 223.0 $\pm$ 0.3 &   &  \\
ONC028    &  C & 358  & 0.13 $\pm$ 0.01 & 8.9 & 05:35:10.121295 & 0.114 & -05:22:32.64849 & 0.191 & 243.7 $\pm$ 0.1 &     & M2 \\
ONC030    &  2 & 363  & 0.16 $\pm$ 0.02 & 9.5 & 05:35:10.252282 & 0.157 & -05:21:57.11306 & 0.269 & 144.0 $\pm$ 0.5 &     & M5.5 \\
ONC032    &  A & 378  & 0.11 $\pm$ 0.01 & 9.6 & 05:35:10.494702 & 0.140 & -05:22:45.75138 & 0.238 & 15.9 $\pm$ 0.1 &   &  \\
ONC035.1  &  A & 390  & 0.11 $\pm$ 0.01 & 14.8 & 05:35:10.597707 & 0.138 & -05:22:55.66422 & 0.261 & 393.9 $\pm$ 0.7 &   &  \\
ONC035.2  &  5 & 390  & 0.28 $\pm$ 0.02 & 13.9 & 05:35:10.615765 & 0.113 & -05:22:56.05438 & 0.160 & 32.5 $\pm$ 0.1 &   &  \\
ONC037    &  3 & 394  & 0.19 $\pm$ 0.01 & 26.1 & 05:35:10.736699 & 0.115 & -05:23:44.72561 & 0.148 & 72.9 $\pm$ 0.1 &     & K2-M0 \\
ONC037    &  7 & 394  & 0.38 $\pm$ 0.02 & 16.4 & 05:35:10.731601 & 0.116 & -05:23:44.66254 & 0.161 & 24.3 $\pm$ 0.1 &     & K2-M0 \\
ONC037    &  8 & 394  & 0.30 $\pm$ 0.01 & 20.5 & 05:35:10.731600 & 0.111 & -05:23:44.66283 & 0.134 & 24.4 $\pm$ 0.1 &     & K2-M0 \\

		\hline
	\end{tabular}
        \begin{tabular}{l}
        NOTE - This table is available in its entirety for download. A portion is shown here. \\
        Columns are (left to right): source number [from \citep{Forbrich:2016}], epoch number, COUP number [from \cite{Getman:2005}], Flux density, \\ J2000 right ascension and declination, both with uncertainties, total separation from the phase centre, YSO class [from \cite{Prisinzano:2008}], \\ spectral type [from \cite{Hillenbrand:2013} and \cite{Skiff:2009}].\\
        The separations are: $\Delta \alpha = (\alpha_{\text{VLBA}} - \alpha_{\text{PC}} \cdot \text{cos} \delta)$, $\Delta \delta = (\delta_{\text{VLBA}} - \delta_{\text{PC}})$, $\Delta \theta = \sqrt{\Delta \alpha^2 + \Delta \delta^2}$ where `VLBA' is the detected source position and \\ `PC' is the respective phase centre. \\
        Epoch numbers are: 1 = BF117, 2 = BF123A, 3 = BF123B, 4 = BF123C, 5 = BF131A, 6 = BF131B, 7 = BF131C, 8 = BF131D, A = BF123AB, \\ B = BF131AB, C = BF131CD \\
        * Source is outside the $\sim$80 per cent power point of the FWHM and the uncertainties might be underestimated.
        \end{tabular}
\end{table}

\end{landscape}

\newpage

\begin{landscape}

\begin{table}
	\centering
	\caption{Table of flux density measurements for the concatenated data in mJy beam$^{-1}$ from the resulting \texttt{imfit} \citep{CASA2022} and the inter-epoch variability factors for each concatenated VLBA data epoch: BF123AB, BF131AB, BF131CD. Variability factors that are listed as being more than (>) are single detected sources and represent lower limits.}
	\label{tab:conc_table}
	\begin{tabular}{lccccc} 
		\hline
        \hline
		[FRM2016] & COUP & BF123AB           & BF131AB           & BF131CD          & Variability Factor\\
		\hline
        021       & 343  & 0.13 $\pm$ 0.01   &  -                & -                & >1.3\\
        025       & -    & 0.10 $\pm$ 0.01   &  -                & -                & >1.0\\
        028       & 358  & -                 &  -                & 0.12 $\pm$ 0.01  & >1.4\\
        032       & 378    & 0.11 $\pm$ 0.01   &  -                & -                & >1.0\\
        035       & 390  & 0.10 $\pm$ 0.01   &  -                & -                & >1.0\\
        047       & -    & 0.08 $\pm$ 0.01   &  -                & -                & >1.0\\
        058       & -    & -                 &  -                & 0.12 $\pm$ 0.01  & >1.4\\
        067       & -    & -                 & 0.13 $\pm$ 0.01   & -                & >1.5\\
        086       & -    & 0.12 $\pm$ 0.01   &  -                & -                & >1.1\\
        116       & -    & -                 &  -                & 0.13 $\pm$ 0.02  & >1.6\\
        128       & -    & -                 & 0.12 $\pm$ 0.01   & -                & >1.4\\
        149       & -    & 0.14 $\pm$ 0.01   &  -                & -                & >1.3\\
        203$^a$       & 655  & 0.12 $\pm$ 0.01   &  -                & -                & 1.6 $\pm$ 0.2\\
        242       & 723  & 0.09 $\pm$ 0.01   &  -                & -                & >1.0\\
        256       & 747  & -                 & 0.13 $\pm$ 0.01   & -                & >1.5\\
        264       & 758  & -                 & 0.14 $\pm$ 0.01   & -                & >1.7\\
        285       & 783  & 0.12 $\pm$ 0.01   &  -                & -                & >1.1\\
        286       & 789    & -                 & 0.14 $\pm$ 0.02   & -                & >1.6\\
        294       & 799    & -                 & 0.12 $\pm$ 0.02   & -                & >1.4\\
        298       & -    & -                 &  -                & 0.15 $\pm$ 0.02  & >1.8\\
        357       & 896  & 0.13 $\pm$ 0.01   &  -                & -                & >1.2\\
        373       & -    & 0.08 $\pm$ 0.01   &  -                & -                & >1.0\\
        401       & 964    & -                 &  -                & 0.12 $\pm$ 0.02  & >1.4\\
        413.1     & -    & -                 & 0.13 $\pm$ 0.02   & -                & >1.5\\
        413.2     & -    & -                 & 0.12 $\pm$ 0.01   & -                & >1.3\\
        420       & -    & -                 & 0.14 $\pm$ 0.02   & -                & >1.6\\
        451       & -    & -                 & 0.16 $\pm$ 0.02   & -                & >1.8\\
        470$^a$       & -    & -                 & 0.12 $\pm$ 0.01   & -                & 1.0$\pm$ 0.1\\
        473       & 1107 & -                 &  -                & 0.15 $\pm$ 0.01  & >1.7\\
        482       & -    & -                 &  -                & 0.14 $\pm$ 0.01  & >1.6\\
        484       & -    & -                 &  -                & 0.13 $\pm$ 0.01  & >1.5\\
        489       & 1143 & -                 &  -                & 0.15 $\pm$ 0.01  & >1.7\\
        514$^a$       & -    & -                 & 0.13 $\pm$ 0.01   & -                & 1.2 $\pm$ 0.1\\
        530$^a$       & 1289    & -              & 0.14 $\pm$ 0.02   & -                & 1.2 $\pm$ 0.1\\

		\hline
	\end{tabular}
    \begin{tabular}{l}
    $^a$ These sources are detected in the normal and concatenated data which then allows the variability factor to be derived. \\
    \end{tabular}
\end{table}

\end{landscape}



\newpage

\bibliographystyle{mnras}
\bibliography{refs} 



\newpage
\appendix


\section{}
\label{appxA}

\subsection{VLBA Sources at Large Separations}

\begin{figure*}
    \includegraphics[width=0.65\paperwidth]{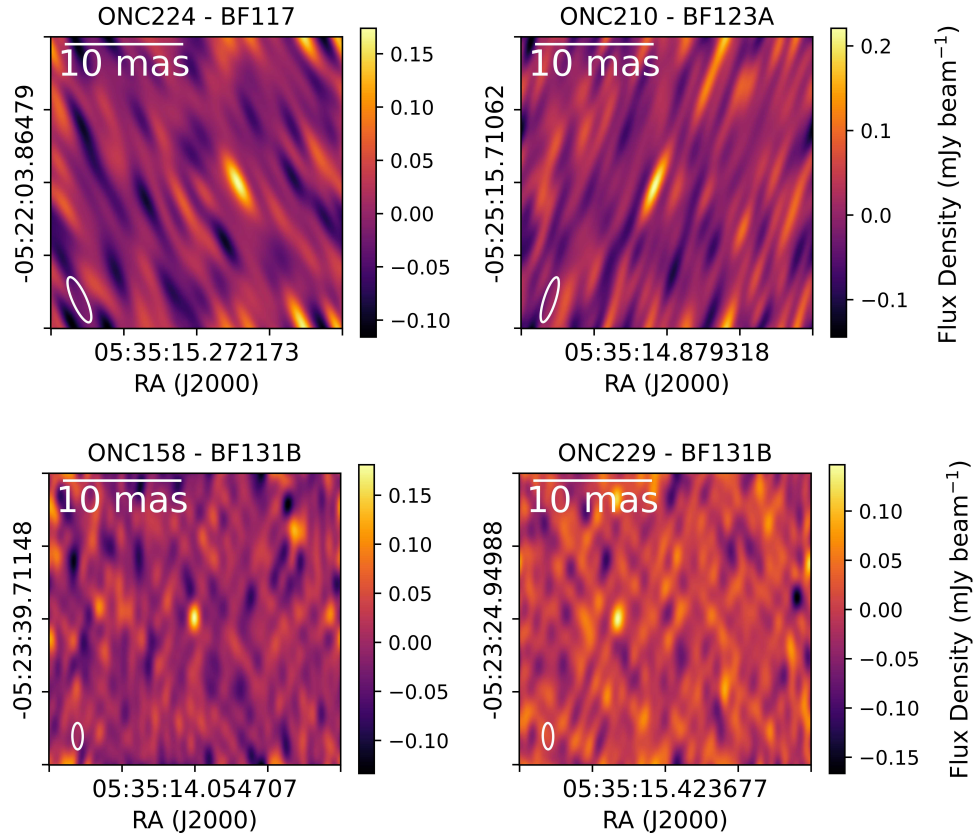}
    \caption{Grid plot showing the images of the four brightest source detection candidates that are at high separations from their respective phase centre. The coordinates of the centre of the image are shown, bottom left is the synthesised beam and upper left is a scale bar showing 10 mas ($\sim$4 AU at the distance of the ONC).}
    \label{fig:EXTRA_grid}
\end{figure*}

We find a subset of sources at larger separations ($>$800 mas) from their respective phase centre. These detections are likely to be genuine, but the effective detection threshold increases with distance from the phase centre. We therefore report them separately from our main sample. Figure\,\ref{fig:EXTRA_grid} shows the oversampled images (see section \ref{sect:data_analysis}) of four of these sources that have the highest signal-to-noise. We search for counterparts for these sources using the \textit{Hubble} catalogue of \cite{Whitmore:2016} which is a master catalogue from archival HST imaging with the ACS, WFPC2 and WFC3 instruments, and the VISTA near-infrared survey of the Orion A molecular cloud called VISION of \cite{Meingast:2016}. We use a search radius of 0.5~arcsec as we did in section \ref{subsect:sp type ev classes}, and find 5 of these sources have counterparts from the \textit{Hubble} catalogue and find no counterparts in the VISION catalogue. The sources that have \textit{Hubble} counterparts are listed in Table \ref{tabApxA:hubble_src_seps} with the separation between the VLBA detection and the counterpart. Additionally, Table \ref{tab:extra_srcs} shows source details for all VLBA sources at larger separations. Where the separation is defined as,

\begin{equation}
    \begin{split}     
    \Delta \alpha & = (\alpha_{\text{VLBA}} - \alpha_{\text{PC}}) \cdot \textrm{cos} \delta \\
    \Delta \delta & = (\delta_{\text{VLBA}} - \delta_{\text{PC}}) \\
    \Delta \theta & = \sqrt{\Delta \alpha^2 + \Delta \delta^2}
    \end{split}
    \label{eq:ang_sep}
\end{equation}

\noindent
where $\Delta \alpha$ is the separation in right-ascension, $\Delta \delta$ is the separation in declination and $\Delta \theta$ is the total separation between the VLBA position and the phase centre (denoted by `PC'). It should be noted that the COUP counterparts listed in Table \ref{tab:extra_srcs} are associated with the respective phase centre and might not be directly related with the VLBA sources due to their large angular separation.

\newpage
\begin{table}
    \centering
    \caption{VLBA sources at larger separations from their respective phase centres that have \textit{Hubble} counterparts and the separations between the VLBA source positions and the \textit{Hubble} counterpart \citep{Whitmore:2016}.}
    \label{tabApxA:hubble_src_seps}
    \begin{tabular}{c|c}
    \hline
    \hline
    [FRM2016] & $\Delta \theta$ (mas) \\
    \hline
    ONC042 & 389.2 $\pm$ 0.2 \\
    ONC224 & 101.7 $\pm$ 0.3 \\
    ONC277 & 362.3 $\pm$ 0.2 \\
    ONC349 & 86.0 $\pm$ 0.2 \\
    ONC484 & 63.7 $\pm$ 0.2 \\
    \hline
    \end{tabular}
\end{table}

\newpage
\begin{landscape}

\begin{table}
	\centering
        \caption{Table of sources at large separations ($>$800 mas) from their respective phase centre.}
        \label{tab:extra_srcs}
	\begin{tabular}{lcccccccc} 
		\hline
        \hline
		[FRM2016] & COUP$^{\text{A}}$ & Flux Density & S/N & $\alpha_{\text{J2000}}$ & $\sigma_{\alpha}$ & $\delta_{\text{J2000}}$ & $\sigma_{\delta}$ & $\Delta \theta$ (mas)$^1$ \\
                  &        & (mJy beam$^{-1}$) &    &                           & (mas) & & (mas) &\\
		\hline
        \textbf{BF117:} \\
        ONC223 & -- & 0.24 $\pm$ 0.02 & 10.6 & 05:35:15.219164 & 0.160 & -05:23:58.59288 & 0.333 & 1066.2 $\pm$ 6.5\\
        ONC224 & -- & 0.19 $\pm$ 0.01 & 13.6 & 05:35:15.272173 & 0.106 & -05:22:03.86479 & 0.213 & 1365.0 $\pm$ 3.8\\
        \hline
        \textbf{BF123A:} \\
        ONC210 & -- & 0.37 $\pm$ 0.02 & 15.7 & 05:35:14.879318 & 0.109 &  -05:25:15.71062 &  0.317 & 979.7 $\pm$ 3.7\\
        \hline
        \textbf{BF123C:} \\
        ONC364 & -- & 0.21 $\pm$ 0.02 & 9.8 & 05:35:17.658061  & 0.021 &  -05:23:39.31981 &  0.116 & 1565.2 $\pm$ 0.3\\
        \hline
        \textbf{BF131A:} \\
        ONC029 & 365 & 0.15 $\pm$ 0.02 & 9.3 & 05:35:10.162020 & 0.046 & -05:23:22.48437 & 0.158 & 1212.9 $\pm$ 0.8\\
        ONC042 & --   & 0.19 $\pm$ 0.01 & 13.2 & 05:35:10.916948 & 0.027 & -05:23:27.58121 & 0.127 & 1016.3 $\pm$ 0.3\\
        ONC054 & --   & 0.14 $\pm$ 0.02 & 8.6 & 05:35:11.512994 & 0.048 & -05:22:55.26907 & 0.149 & 1070.4 $\pm$ 0.7\\
        ONC077 & 470 & 0.16 $\pm$ 0.02 & 8.7 & 05:35:12.271003 & 0.083 & -05:23:48.93483 & 0.246 & 894.9 $\pm$ 1.6\\
        ONC532* & 1296 & 0.49 $\pm$ 0.05 & 11.5 & 05:35:24.573942 & 0.059 & -05:19:32.34703 & 0.151 & 907.7 $\pm$ 0.9\\
        \hline
        \textbf{BF131B:} \\
        ONC072 & -- & 0.16 $\pm$ 0.01 & 11.4 & 05:35:12.119197 & 0.024 & -05:22:13.17631 & 0.147 & 1015.5 $\pm$ 0.3\\
        ONC120 & -- & 0.19 $\pm$ 0.02 & 11.8 & 05:35:13.310415 & 0.019 & -05:23:28.71650 & 0.147 & 903.7 $\pm$ 0.2\\
        ONC158 & 602 & 0.23 $\pm$ 0.02 & 13.7 & 05:35:14.054707 & 0.014 & -05:23:39.71148 & 0.101 & 1272.1 $\pm$ 0.1\\
        ONC229 & -- & 0.19 $\pm$ 0.01 & 19.5 & 05:35:15.423677 & 0.017 & -05:23:24.94988 & 0.069 & 1025.7 $\pm$ 0.1\\
        ONC349 & 879 & 0.14 $\pm$ 0.01 & 9.5 & 05:35:17.390507 & 0.029 & -05:23:05.86332 & 0.187 & 996.9 $\pm$ 0.4\\
        ONC409 & 976 & 0.24 $\pm$ 0.03 & 10.4 & 05:35:18.480052 & 0.051 & -05:20:43.60629 & 0.160 & 962.6 $\pm$ 0.7\\
        ONC484 & -- & 0.20 $\pm$ 0.02 & 11.1 & 05:35:21.070363 & 0.027 & -05:22:26.14951 & 0.141 & 956.0 $\pm$ 0.3\\
        \hline
        \textbf{BF131C:} \\
        ONC240 & 717 & 0.31 $\pm$ 0.02 & 11.5 & 05:35:15.573475 & 0.037 & -05:23:38.08918 & 0.139 & 865.0 $\pm$ 0.4\\
        \hline
        \textbf{BF131D:} \\
        ONC210 & -- & 0.26 $\pm$ 0.02 & 10.8 & 05:35:14.727479 & 0.025 & -05:25:13.89417 & 0.182 & 1583.0 $\pm$ 0.5\\
        ONC277 & 780 & 0.12 $\pm$ 0.01 & 10.9 & 05:35:16.134949 & 0.017 & -05:23:24.06439 & 0.165 & 1080.8 $\pm$ 0.3\\
        \hline
	\end{tabular}
    \begin{tabular}{l}           
        Columns are (left to right): source number from \citep{Forbrich:2016}, COUP number from \citep{Getman:2005}, flux density, \\ signal-to-noise, J2000 right ascension and declination, both with uncertainties, total separation from nominal VLA position. \\
        * Source is outside the $\sim$80 per cent power point of the FWHM and the uncertainties might be underestimated. \\
        $^{\text{A}}$ The respective phase centre is associated with a COUP counterpart, as these sources are at high separations \\ they might not be directly associated with the X-ray counterpart.
        \end{tabular}
\end{table}

\end{landscape}


\section{}
\label{appxB}

This appendix discusses more details of specific sources that exhibit higher variability with VFs $\geq$ 5 (e.g., COUP number, mass estimates, spectral types from \cite{Hillenbrand:2013} and/or YSO class from \cite{Prisinzano:2008} where available).

\subsubsection{ONC003.1}
ONC003.1 is only detected in BF131C with a peak flux density of 0.71 $\pm$ 0.08 mJy beam$^{-1}$. A lower limit for the VF of this source is >6.1 using the the 5.5$\sigma$ value of the lowest rms as an upper limit, see Section \ref{subsect:VF} for discussion of variability factors. Unlike the other higher variable sources (discussed below) ONC003 does not have a COUP counterpart. As detailed in section \ref{sect:results_discussion} this could be due to variability or this source could be a very deeply embedded object that is not detected in X-ray. This source does have a VLBA radio counterpart (ONC003.2 detected in BF131D) with a separation of 0.50 arcsec, see Table \ref{tab:source_detections}.

\subsubsection{ONC016 (COUP 322 or 2MASS J05350873-0522566)}
\cite{Hillenbrand:2013} lists ONC016 as spectral type M1.5. \cite{Wei:2024} estimates the mass using the `MIST' stellar evolutionary model as 0.829 $\pm$ 0.064 $M_\odot$ as part of their near-infrared `NIRSPEC' work using the Keck II 10 m telescope. \cite{Getman_Feig:2021} estimate the mass as 0.5 $M_{\odot}$ from their \textit{Chandra} study. This source is detected in BF131A, as seen in Table \ref{tab:flux_table}, with an averaged peak flux density of 1.56 $\pm$ 0.03\,mJy\,beam$^{-1}$ for the whole epoch. As it is a single detection a lower limit of the VF is >12, this can be seen in Fig. \ref{fig:VLBA_epoch_LC}.

\subsubsection{ONC066 (COUP 450 or GMR A)}
ONC066 is detected in all VLBA epochs as seen in Table \ref{tab:flux_table}. It has an averaged peak flux density of 10.27 $\pm$ 0.18 mJy beam$^{-1}$ in BF131C and lowest flux density of 1.16 $\pm$ 0.02 mJy beam$^{-1}$ in BF123B, this corresponds to a highest variability factor of 8.8 $\pm$ 0.2. The increase of flux density in BF131C also results in the variability factor being higher between BF131B and BF131C at VF = 5.0 $\pm$ 0.2 (representing a time-scale of $\sim$1 year) and between BF131C and BF131D (representing a time-scale of 1 day) at VF = 3.1 $\pm$ 0.1. These higher VFs point to potential variability occurring for ONC066 on shorter time-scales at least in the case of BF131C and the adjacent epochs.

\subsubsection{ONC154 (COUP 594)}
\cite{Zapata:2004-2} concluded that this is a counterpart of the X-ray source COUP 594 as part of their VLA survey, and detections of circular polarisation ($\sim$20 per cent right-circular polarisation) indicates gyrosynchrotron emission associated with a pre-main sequence star. Table \ref{tab:flux_table} shows the highest averaged peak flux density is 1.91 $\pm$ 0.04 mJy beam$^{-1}$ in BF123C and a lowest flux density of 0.17 $\pm$ 0.02 mJy beam$^{-1}$ in BF131D, this corresponds to VF = 12.0 $\pm$ 1.8.

\subsubsection{ONC196 (COUP 648)}
ONC196 is classified by \cite{Hillenbrand:2013} as spectral type K3-M2, and \cite{Getman_Feig:2021} estimate the mass as 2.5 $M_{\odot}$ from their \textit{Chandra} study. It is detected in BF117 epoch and has a peak flux density of 0.69 $\pm$ 0.03 mJy beam$^{-1}$. As this source is only detected once a lower limit of the variability factor was estimated as >6.0.

\subsubsection{ONC212 (COUP 670)}
ONC212 is classified as spectral type K3-M2 by \cite{Hillenbrand:2013}. The mass is listed as 0.434 $\pm$ 0.032 $M_\odot$ by \cite{Wei:2024} from their `MIST' model using near-infrared `NIRSPEC' data from the Keck II 10 m telescope. The peak flux density is 1.06 $\pm$ 0.02 mJy beam$^{-1}$. This source is only detected in BF117 epoch, therefore, the lower limit of the VF is >9.2.

\subsubsection{ONC219 (COUP 680)}
ONC219 is only detected in BF131D epoch, with a peak flux density of 0.62 $\pm$ 0.03 mJy beam$^{-1}$, so a lower limit of the VF was calculated as >5.3. This source is also discussed by \cite{Vargas-gonz:2023} as a millimetre flare source observed with the Atacama Large Millimetre/submillimetre Array, there are 9 other sources discussed by \cite{Vargas-gonz:2023} that have centimetre and millimetre counterparts.

\subsubsection{ONC241 (COUP 718 or V1501 Ori)}
\cite{Morales-Calderon:2012} classified ONC241 as an accreting T-Tauri star and possible eclipsing binary as part of their Spitzer Exploration Science Program YSOVAR, and \cite{Sicilia-Aguilar:2005} concluded it is likely a weak-lined T-Tauri star from their 6.5 m MMT data. Additionally, \cite{Hillenbrand:2013} gives a spectral classification of K4-M1. \cite{Wei:2024} estimates the mass using the `MIST' \citep{Choi:2016, Dotter:2016} stellar evolutionary model as 0.892 $\pm$ 0.063 $M_\odot$ using near-infrared `NIRSPEC' data from the Keck II 10 m telescope. Figure\,\ref{fig:VLBA_epoch_LC} and Table \ref{tab:flux_table} shows the highest peak averaged flux density is 2.99 $\pm$ 0.07 mJy beam$^{-1}$ in BF131A and the lowest is 0.22 $\pm$ 0.02 in the BF123B epoch which corresponds to VF = 13.5 $\pm$ 1.2 on a time-scale of 1 year.

\subsubsection{ONC300.2 (COUP 801)}
ONC300.2 is classified as spectral type K4-K7 by \cite{Hillenbrand:2013}. The mass is listed as 0.437 $\pm$ 0.031 $M_\odot$ by \cite{Wei:2024} from the `MIST' model \citep{Choi:2016, Dotter:2016}. The peak flux density is 0.97 $\pm$ 0.04 mJy beam$^{-1}$ in BF131B, resulting in a lower limit of >8.2 for the VF. This source has a VLBA radio counterpart (ONC300.1 detected in BF117) with a separation of 0.57 arcsec, see Table \ref{tab:source_detections}.

\subsubsection{ONC400 (COUP 965 or V1229 Ori)}
ONC400 is classified in the The General Catalogue of Variable Stars (GCVS) as an irregular eruptive variable of type 'IN' or Orion variable \citep{Samus:2017} and as spectral type G8-M2 by \cite{Hillenbrand:2013}. \cite{Garay:1987} discusses this source as a T-Tauri star that exhibits radio variability from their VLA observations at 1.5, 4.9, 15.0, and 22.5 GHz and \cite{Sheehan:2016} observed this with the VLA as a radio variable star at 3.6 cm (8 GHz). The mass was estimated by \cite{Wei:2024} as 0.487 $\pm$ 0.063 $M_\odot$ using the `MIST' stellar model \citep{Choi:2016, Dotter:2016} using near-infrared `NIRSPEC' data from the Keck II 10 m telescope. ONC400 has a peak averaged flux density of 14.38 $\pm$ 0.32 mJy beam$^{-1}$ in BF123C and lowest of 2.10 $\pm$ 0.06 in BF123A which corresponds to VF = 6.8 $\pm$ 0.3 on a time-scale of 1 yr.

\subsubsection{ONC414.1 (COUP 985 or MV Ori)}
ONC414.1 is classified as spectral type F8-K4 by \citep{Hillenbrand:2013}. It is a radio variable source as discussed by \cite{Forbrich:2017}. Table \ref{tab:flux_table} shows it has a peak averaged flux density of 0.69 $\pm$ 0.07 mJy beam$^{-1}$ in BF117, and a lowest flux density of 0.15 $\pm$ 0.01 mJy beam$^{-1}$ in BF123C which results in VF = 4.5 $\pm$ 0.6. This source has a counterpart (ONC414.2), also discussed by \cite{Dzib2021}, detected in epochs BF123B and BF123C at a separation of 0.22 arcsec, see Table \ref{tab:source_detections}.

\subsubsection{ONC444 (COUP 1035)}
ONC444.1 and ONC444.2 are two separate sources detected in BF131B and BF131D as seen in Table \ref{tab:source_detections} where the separation from the nominal VLA position is: 502.08 $\pm$ 1.30 and 133.98 $\pm$ 0.18 respectively. Considering the time-scale between the detections of only a year and the large separation these are likely two separate sources. \cite{Megeath:2012} classifies ONC444.2 as a YSO with a disk using mid-infrared data from Spitzer. Their peak flux densities are: ONC444.1 = 3.76 $\pm$ 0.43 mJy beam$^{-1}$ resulting in a lower limit of >32.4 for the VF and ONC444.2 = 3.57 $\pm$ 0.25 mJy beam$^{-1}$ which gives a lower limit of >30.8 for the VF. It should be noted these sources sit outside the $\sim$80 per cent power point of the FWHM where they are both $\sim5.5'$ from the pointing centre so the uncertainties could be underestimated, these are flagged in Tables \ref{tab:flux_table} and \ref{tab:source_detections} to show this.

\subsubsection{ONC459 (COUP 1083 or V* V1517 Ori)}
\begin{landscape}
    
\begin{figure}
    \centering
    \includegraphics[width=0.8\paperwidth]{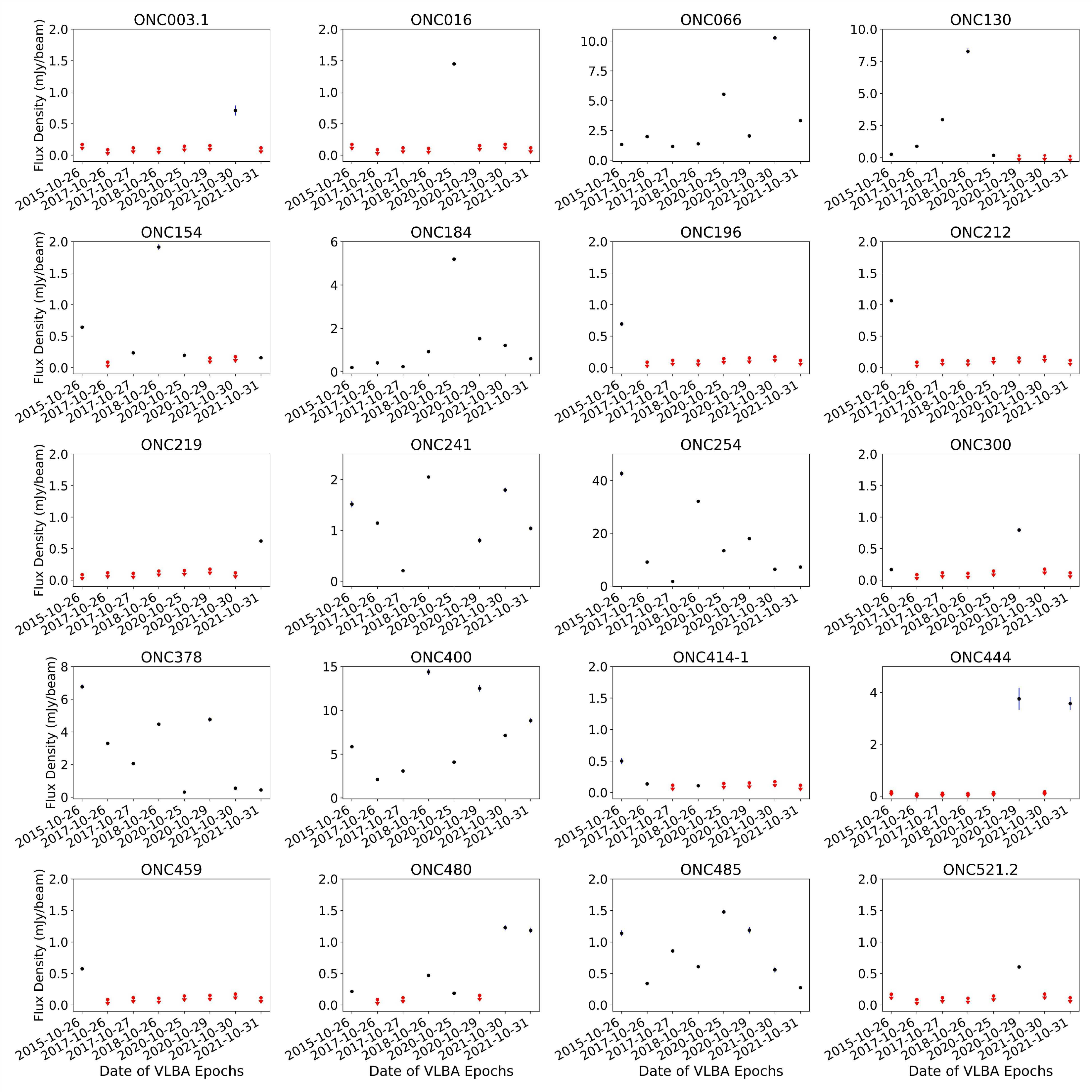}
    \caption{Grid plot of light curves for sources with higher variability factors ($\geq 5$). This plot shows the primary beam corrected flux density measurements for the source for each epoch as measured by `imfit'. The flux density is shown in black, and where a source was not detected the upper limit of the 5.5$\sigma$ value of the rms is shown in red.}
    \label{fig:VLBA_epoch_LC}
\end{figure}

\end{landscape}

ONC459 is classified as a class III YSO by \cite{Prisinzano:2008} and \cite{Hillenbrand:2013} classifies it as spectral type <M0. It is only detected in BF117 epoch with a flux density of 0.65 $\pm$ 0.02 mJy beam$^{-1}$. Therefore, a lower limit of the VF was calculated as >4.9 using the 5.5$\sigma$ value of the lowest rms value which can be seen in Fig. \ref{fig:VLBA_epoch_LC}.

\subsubsection{ONC480 (COUP 1116 or V1230 Ori)}
ONC480 is classified as spectral type B by \cite{Hillenbrand:2013}. \cite{Sheehan:2016} found this is a variable source at 1.3 cm (23 GHz) from their VLA observations. The peak averaged flux density is 1.23 $\pm$ 0.04 in BF131C and lowest of 0.19 $\pm$ 0.02 in BF131A, this results in VF = 6.5 $\pm$ 0.4 across a timespan of 1 yr. \cite{Dzib:2026b} discuss this source and conclude from their astrometry that it is likely a binary system with a low-mass YSO companion. They suggested that the primary star, of spectral type B1, is the source detected by \textit{Gaia}, while the companion is likely a lower-mass star responsible for the non-thermal emission detected here.

\subsubsection{ONC485 (COUP 1130 or V1399 Ori)}
ONC485 is classified as spectral type G8-K5 by \cite{Prisinzano:2008}. \cite{Wei:2024} estimates the mass using the `MIST' stellar evolutionary model as 0.446 $\pm$ 0.027 $M_\odot$ from their near-infrared `NIRSPEC' work from Keck II 10 m telescope observations. The peak averaged flux density is 1.48 $\pm$ 0.03 mJy beam$^{-1}$ in BF131A and the lowest is 0.27 $\pm$ 0.02 mJy beam$^{-1}$ in BF131D. This corresponds to VF = 5.5 $\pm$ 0.5 across a timespan of 1 year. 

\subsubsection{ONC521.2 (COUP 1261 or V* V1342)}
ONC521.2 is detected in BF131B at a peak flux density of 0.60 $\pm$ 0.02 mJy beam$^{-1}$. It is spectral type M3-M3.5e \citep{Hillenbrand:2013} and has a lower limit of >5.2 for the VF. This source has a VLBA counterpart detected in epoch BF117 at a separation of 0.31, see Table \ref{tab:source_detections}. This source is outside the $\sim$80 per cent power point of the FWHM at a separation of $\sim3.3$ arcmin from the pointing centre. This source is flagged in Tables \ref{tab:flux_table} and \ref{tab:source_detections} as having potentially underestimated uncertainties.

\bsp	
\label{lastpage}
\end{document}